\documentclass{aa}

\usepackage{graphicx}
\usepackage{txfonts}
\usepackage{hyperref}
\usepackage{longtable}
\usepackage{isotope}

\newcommand{\msun}{$\rm M_{\odot}$}
\newcommand{\p}{$p$}

\newcommand{\g}{$\gamma$}

\newcommand{\mco}{$\rm M_{CO}$}
\newcommand{\xc}{$\rm X_{C12}$}
\newcommand{\cago}{\isotope[12]{C}($\alpha$,$\gamma$)\isotope[16]{O}}

\begin{document}

\title{The Occurrence and Impact of Carbon-Oxygen Shell Mergers in Massive Stars}

\author{
L. Roberti\inst{1,2,3,4,5} 
\and M. Pignatari\inst{2,3,6,5} 
\and H. E. Brinkman\inst{7,5} 
\and S. K. Jeena\inst{8} 
\and A. Sieverding\inst{9} 
\and A. Falla\inst{10,4}
\and M. Limongi\inst{4,11,12} 
\and A. Chieffi\inst{13,14,12} 
\and M. Lugaro\inst{2,3,15,14} 
}

\institute{
Istituto Nazionale di Fisica Nucleare - Laboratori Nazionali del Sud, Via Santa Sofia 62, Catania, I-95123, Italy
\and Konkoly Observatory, Research Centre for Astronomy and Earth Sciences, HUN-REN, Konkoly Thege Miklós út 15-17, Budapest, H-1121, Hungary 
\and CSFK HUN-REN, MTA Centre of Excellence, Konkoly Thege Miklós út 15-17, Budapest, H-1121, Hungary
\and Istituto Nazionale di Astrofisica – Osservatorio Astronomico di Roma, Via Frascati 33, Monte Porzio Catone, I-00040, Italy
\and NuGrid Collaboration, \url{http://nugridstars.org}
\and E. A. Milne Centre for Astrophysics, University of Hull, Cottingham Road, Kingston upon Hull, HU6 7RX, UK
\and Institute of Astronomy, KU Leuven, Celestijnenlaan 200D, 3001, Leuven, Belgium
\and Department of Physics, Indian Institute of Technology Palakkad, Kerala, India
\and Lawrence Livermore National Laboratory, 7000 East Ave, Livermore, CA, 94550, USA
\and Dipartimento di Fisica, Sapienza Università di Roma, P.le A. Moro 5, Roma, I-00185, Italy
\and Kavli Institute for the Physics and Mathematics of the Universe, Todai Institutes for Advanced Study, University of Tokyo, Kashiwa, 277-8583 (Kavli IPMU, WPI), Japan
\and Istituto Nazionale di Fisica Nucleare, Sezione di Perugia, via A. Pascoli s/n, Perugia, I-06125, Italy
\and Istituto Nazionale di Astrofisica—Istituto di Astrofisica e Planetologia Spaziali, Via Fosso del Cavaliere 100, Roma, I-00133, Italy
\and School of Physics and Astronomy, Monash University, VIC 3800, Australia
\and ELTE Eötvös Loránd University, Institute of Physics and Astronomy, Budapest 1117, Pázmány Péter sétány 1/A, Hungary
}

\date{}

\abstract
{In their final stages before undergoing a core-collapse supernova, massive stars may experience mergers between internal shells where carbon (C) and oxygen (O) are consumed as fuels for nuclear burning. This interaction, known as a C-O shell merger, can dramatically alter the internal structure of the star, leading to peculiar nucleosynthesis and potentially influencing the supernova explosion and the propagation of the subsequent supernova shock.}  
{Our understanding of the frequency and consequences of C-O shell mergers remains limited. This study aims to identify for the first time early diagnostics in the stellar structure which will lead to C-O shell mergers in more advanced stages. We also assess their role in shaping the chemical abundances in the most metal poor stars of the Galaxy.}  
{We analyze a set of 209 of stellar evolution models available in the literature, with different initial progenitor masses and metallicities. We then compare the nucleosynthetic yields from a subset of these models with the abundances of odd-Z elements in metal-poor stars.}  
{We find that the occurrence of C-O shell mergers in stellar models can be predicted with good approximation based on the outcomes of the central He burning phase, specifically, from the CO core mass (\mco) and the \isotope[12]{C} central mass fraction (\xc): 90$\%$ of models with a C-O merger have \xc $< 0.277$ and \mco $< 4.90$ \msun, with average values \mco = 4.02 \msun\ and \xc= 0.176.
The quantities \xc\ and \mco\ are indirectly affected from several stellar properties, including the initial stellar mass and metallicity.
Additionally, we confirm that the Sc-rich and K-rich yields from models with C-O mergers would solve the long-standing underproduction of these elements in massive stars.
}  
{Our results emphasize the crucial role of C-O shell mergers in enriching the interstellar medium, particularly in the production of odd-Z elements. This highlights the necessity of further investigations to refine their influence on pre-supernova stellar properties and their broader impact on galactic chemical evolution.}  


\maketitle

\section{Introduction} \label{sec:intro}

    Massive stars are responsible for the production of roughly half of the solar abundances of the elements in the periodic table \citep[see, e.g.,][]{prantzos:18,kobayashi:20a}. During their lives, they transform their initial chemical composition, mostly dominated by H and He, into heavier and heavier nuclear species by means of nuclear reactions, until a compact Fe core forms surrounded by layers of materials with variable chemical composition. Massive stars evolve by alternating contraction phases of the core, in which the temperature rises, to central burning phases, where an element is converted into a heavier species. Once a given fuel is exhausted in the center of the star, the contraction resumes, the burning of that fuel shifts outwards in mass (shell burning), and a heavier fuel is ignited in the center \citep[e.g.,][]{woosley:02}. During the spectacular end of their lives as core-collapse supernovae (CCSNe) the products of the hydrostatic evolution contained in these layers are partly reprocessed by explosive nucleosynthesis, therefore, the CCSN ejecta contain the contribution from the nuclear burning that occurred both during the pre-supernova evolution and during the explosion itself \citep{rauscher:02,pignatari:16,LC18,RLC24,BR24}. 

    In the last days of their lives, massive stars can experience the ingestion of carbon (C) and neon (Ne) into the oxygen (O) burning shell, where convection, in the form of turbulent motions, vigorously mixes out the burning products and bring in further fuel \citep[][]{rauscher:02,meakin:06,ritter:18,andrassy:20,rizzuti:24}. This convective-reactive event is referred to as a "C-O shell merger" and can result in the formation of a large mixed convective region, where C-rich material is exposed to the high temperatures typical of O burning. The resulting peculiar nucleosynthetic signature of this merger has been shown to survive reprocessing by explosive nucleosynthesis \citep{rauscher:02,CL17,ritter:18a,roberti:23,roberti:24}. The variety of products from the C-O shell merger are thus ejected into the stellar surrounding, contributing to the chemical enrichment of the interstellar medium. 
    Furthermore, this peculiar merger event may create asymmetries in the stellar structure of the supernova progenitor and cause a significant drop in density at the interface between the Si and O layers. These effects could potentially facilitate the CCSN explosion and influence the oscillation modes of gravitational waves emitted during the explosion \citep[see, e.g.,][]{couch:13,mueller:16a,boccioli:23,torres-forne:19,zha:24,laplace:25}.

    C-O shell mergers have been found in theoretical, spherically symmetric (1D) stellar models, independently from the code used to produce the simulations, generally in massive stars with an initial mass $12 \leq \rm M_{ini} (M_{\odot}) \leq 25$. Their occurrence is also supported by the rather few multidimensional simulations of the final stages of the evolution of massive stars \citep[see, e.g.,][]{meakin:06,mocak:18,andrassy:20,yadav:20}. For example, \cite{rizzuti:22,rizzuti:23} showed that O and Ne convective shells in a 3D model naturally tend to entrain some material from the layers at the edges of the convective zone. This process facilitates the ingestion of material from the adjacent burning shell and may trigger a convective-reactive event \citep{rizzuti:24}. However, direct, or even indirect, observational evidence of the existence in nature of C-O shell mergers is still lacking. \cite{ritter:18a} suggested that a fraction of massive stars in a stellar population is required to undergo C-O mergers to reproduce the galactic abundances of the odd-Z elements, such as P, K, and Sc. However, this analysis was limited to a parametric study based on the ejecta of a single non-rotating 15 \msun\ stellar model of solar metallicity.

    Here we present a study which provides us with a new method to predict the occurrence of C-O shell merger in 1D stellar models. Furthermore, we demonstrate that observations of old stars from the early Universe show the effect of the C-O mergers that occurred in the first generation of massive stars.

\section{Conditions for C-O mergers} \label{sec:c12}

    The advanced burning stages of a massive star are controlled by the extension in mass of the He-exhausted core, rich in C and O (the CO core mass, or \mco), and the \isotope[12]{C} mass fraction left in the center of the star (\xc) after the central He burning phase \citep[][]{CL13,LC18}. The CO core mass has the same role to the advanced burning phases as the initial mass to the main sequence of the star. Since all the following  burning phases take place within the CO core, its size (both in mass and in radius) determines the region available for the development and evolution of the various C, Ne, O, and Si burning shells. In a smaller CO core, these burning shells are closer to each other than that in a larger CO core. The \isotope[12]{C} abundance controls the amount of fuel available for core C burning and, more importantly, the number and the extent of the C burning shells, that contribute to sculpt the final density and entropy profiles at the moment of the core-collapse. In particular, a lower \isotope[12]{C} abundance left by He burning results in a lower \isotope[20]{Ne} abundance
    \footnote{Note that \isotope[20]{Ne} is the main product of C burning via \isotope[12]{C}(\isotope[12]{C},$\alpha$)\isotope[20]{Ne}, \isotope[12]{C}(\isotope[12]{C},$p$)\isotope[23]{Na}($p$,$\alpha$)\isotope[20]{Ne}, and \isotope[16]{O}($\alpha$,$\gamma$)\isotope[20]{Ne}.} 
    and, eventually, in a smaller entropy discontinuity at the interfaces between the O, Ne, and C convective shells \citep{roberti:24}. In general, an entropy discontinuity acts as a barrier sustained by nuclear burning which prevents the underlying shell from penetrating into the overlying one. As the C (and Ne) abundance decreases, the strength of this barrier decreases accordingly. 
    It follows that a small CO core together with a low \isotope[12]{C} abundance at the end of He burning would, in principle, favor the penetration of the O burning region into the C-rich layers. 

    \begin{figure*}[!t]
        \includegraphics[width=1.00\textwidth]{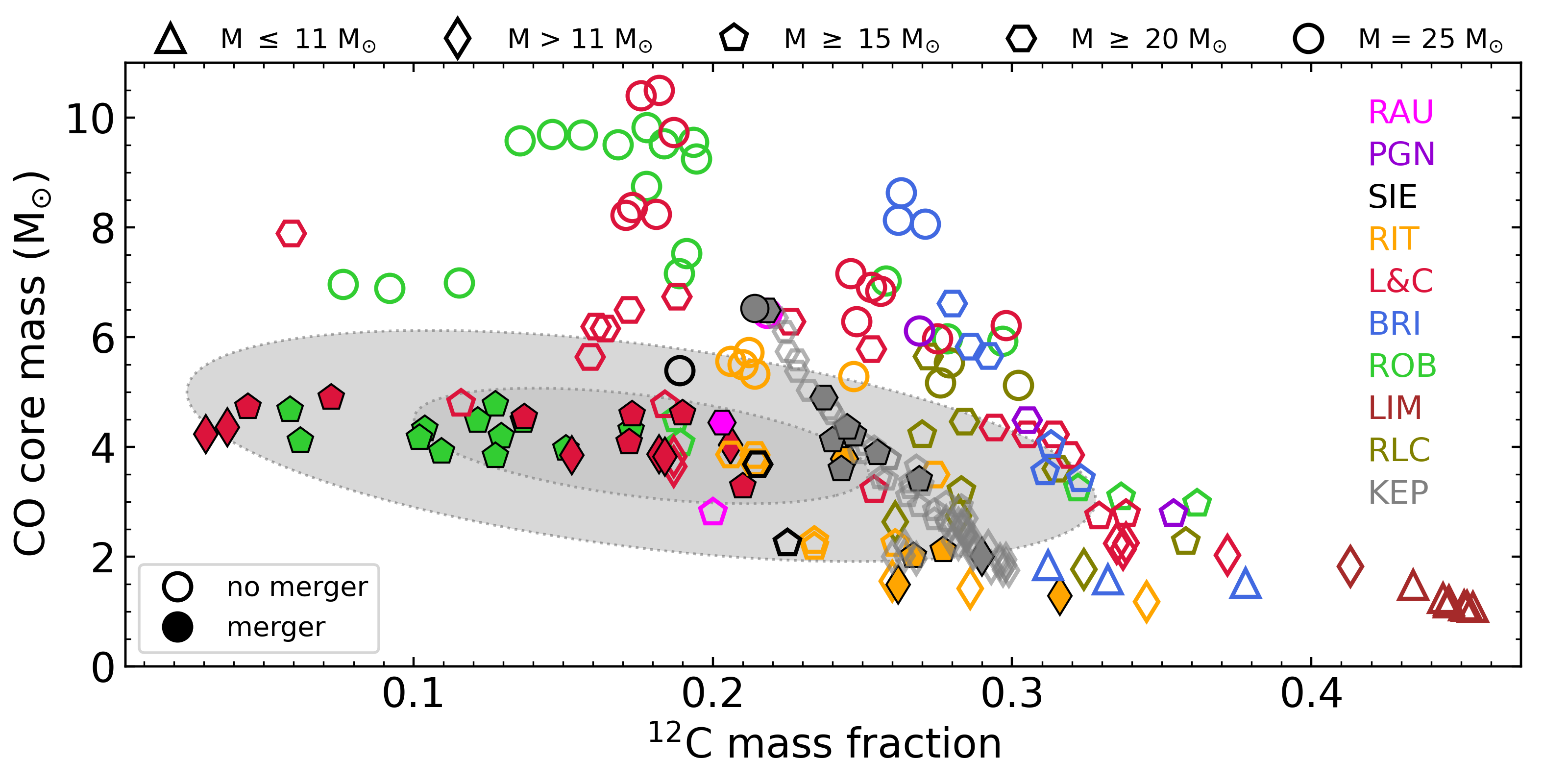}
        \caption{The CO core mass (\mco) versus the central \isotope[12]{C} mass fraction abundance (\xc) at the end of central He burning phase. The different colors identify the different set of models (see side legend) where RAU \cite{rauscher:02}; PGN \cite{pignatari:16}; SIE \cite{sieverding:18}; RIT \cite{ritter:18}; L$\&$C \cite{LC18}; BRI \cite{brinkman:19,brinkman:21}; LIM \cite{limongi:24}; RLC \cite{RLC24}; ROB \cite{roberti:24}; KEP \cite{jeena:24}. Different symbols identify different initial masses (see top legend). Models with and without a C-O shell merger are represented by filled and empty symbols, respectively. The two gray shaded area represent the covariant confidence ellipses relative to the models with C-O mergers within 1 and 2 standard deviations $\sigma$, centred at \mco=4.02 and \xc=0.176.}\label{fig:1}
    \end{figure*}   

    To systematically study this prediction on the occurrence of C-O shell mergers in theoretical calculations, we collected 209 published 1D models with initial masses between 9.2 and 25 \msun, initial metallicity between 0 and 0.05, and initial rotation velocity between 0 and 800 km $\rm s^{-1}$ \citep[][see Sect.\ref{subapp:fig1} for a brief overview of the main properties of the sets of models considered]{rauscher:02,pignatari:16,sieverding:18,ritter:18,LC18,brinkman:19,brinkman:21,limongi:24,jeena:24,RLC24,roberti:24}, and present their \mco\ and \xc\ in \figurename~\ref{fig:1}.  
    Out of 209 models, 41 present a C-O shell merger. While this number constitutes a representative percentage of massive star models with a C-O merger in literature, it is not necessarily representative yet of a real stellar population because it is not weighted over an initial mass function, nor over an initial rotation velocity and metallicity distributions. Significantly, we find that the majority of the models with a C-O merger fall in an identifiable region of the \mco-\xc\ diagram, the grey area in \figurename~\ref{fig:1}. Most of the models without a C-O shell merger are not located in this area (77$\%$, 132 out of 172). Some exceptions are present, which is not surprising and mostly due to the large uncertainty introduced by the adoption of a different physics input (e.g., reaction rates, opacity, mass loss prescriptions) and mixing treatments (e.g., convection and semi-convection parameters, definition of convective boundary mixing, rotational mixing) in every set of models, which can affect the exact parameter space for C-O shell merger events. For example, using the Schwarzschild criterion to define the convective O shell boundary permits a deeper penetration into the Ne-C region, whereas the Ledoux criterion would limit the extent of the convective zone at the chemical discontinuity \cite{clayton:68,kippenhahn:90}. As another example, the adoption of a larger \cago\ rate would result in a lower \xc, but approximately the same \mco. The quantitative impact of these uncertainties can only be assessed by running large grids of models, each varying a single input parameter, all approaches extremely costly both in terms of time and resources.
    
    Figure~\ref{fig:1} demonstrates that the occurrence of a C-O shell merger depends on the outcomes of the central He burning phase (specifically, \mco\ and \xc), and only indirectly on initial mass and metallicity. 
    We find that 90$\%$ of models that undergo a C-O merger have \xc $< 0.277$ and \mco $< 4.90$ \msun, with average values \mco = 4.02 \msun\ and \xc= 0.176. Within this parameter space, we identify 79 models in total, of which 38 do not exhibit a C-O merger. 
    This indicates that the threshold values for \xc\ and \mco\ are a necessary but not sufficient condition to determine whether the merger occurs and that, even when the CO core mass and central carbon abundance fall within the expected range, the likelihood of a merger remains around $50\%$, suggesting that additional mechanisms may play a role in triggering the merger.
    
    While both rotating and non-rotating models are included in the figure, this is not a criterium for the merger. In fact, rotationally induced secular instabilities do not have enough time to affect the latest stages of evolution. Consequently, because of the timescales, the rotation does not influence the mixing coefficient at the interface of the two shells and thus does not directly cause the shell interaction, although it affects it indirectly via changing \mco\ and \xc\ \citep[e.g.,][]{CL13}, similarly to other stellar parameters such as the initial mass and metallicity.

\section{Nuclear Reactions and Nucleosynthesis in a C-O merger}

    \begin{figure}[!t]
        \includegraphics[width=0.48\textwidth]{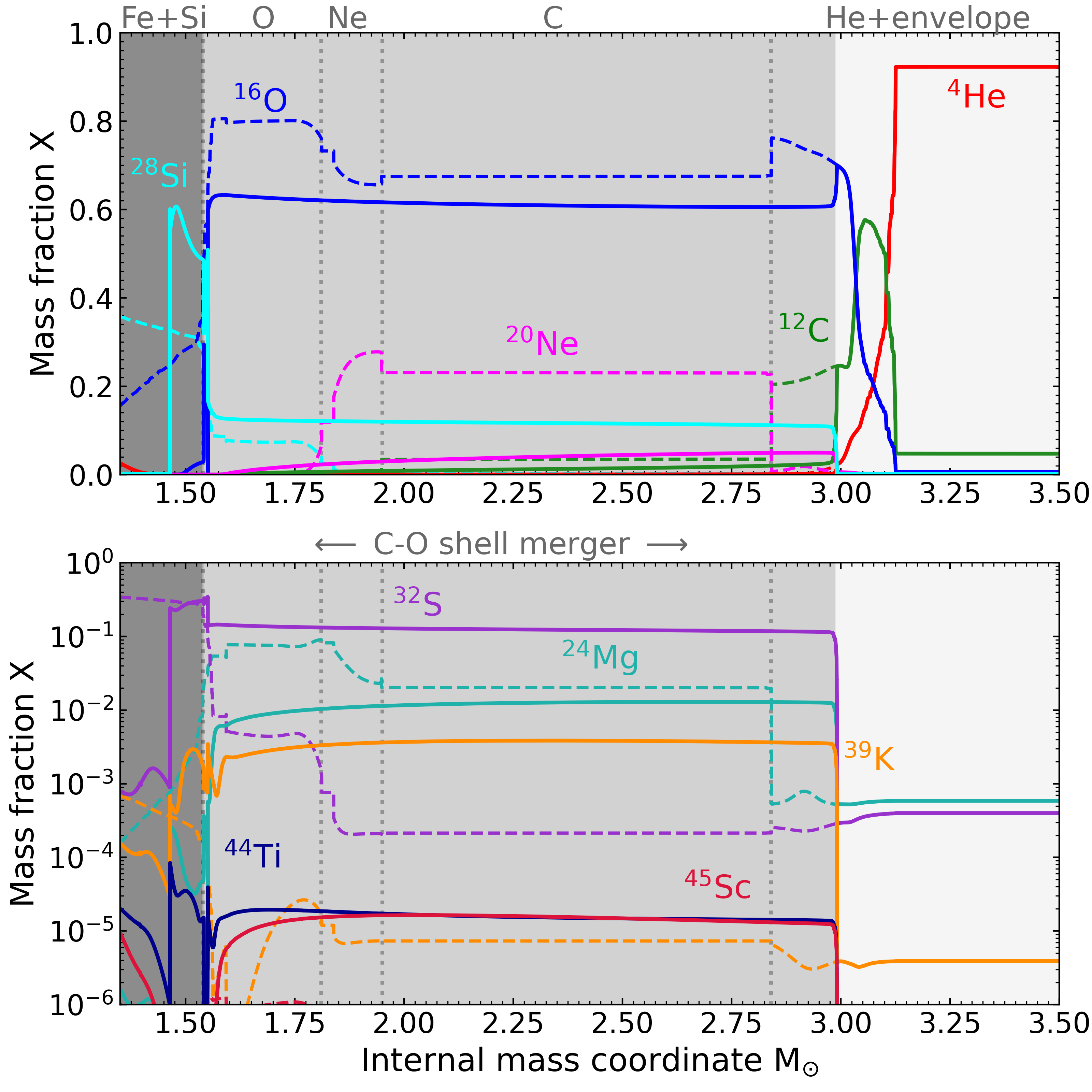}
        \caption{Comparison between the abundances before (dashed lines) and after the C-O shell merger (solid lines) in the RIT 15 \msun\ model at solar metallicity \citep[see][for a detailed description of the model]{ritter:18}. The upper panel shows the effect of the merger on the major fuels. The lower panels show instead the effect of the merger on some key C- and O-burning products and selected odd-Z isotopes (see text). The vertical dotted lines represent the location of the edges of the shells before the C-O shell merger. The light grey area is the region modified by the merger, the dark grey area represents instead the untouched deeper layers of the star after the merger, i.e., the Fe core and the Si shell.}\label{fig:x}
    \end{figure}  
    
    During a C-O shell merger, C-burning products are transported to the base of the O burning shell, where they interact with the products of O-burning at temperatures significantly higher than those typical of C-burning. This interaction leads to a slight reduction in the final abundances of the C-burning products (such as \isotope[23]{Na}, \isotope[24]{Mg}, and \isotope[27]{Al}) and an enhancement of those of the O-burning products (such as \isotope[28]{Si}, \isotope[31]{P}, and \isotope[32]{S}). \figurename~\ref{fig:x} shows this occurrence in the case of a 15 \msun\ star at solar metallicity. The injection of \isotope[12]{C} and fresh \isotope[16]{O} at approximately 2.6 - 2.8 GK (typical temperature of O shell burning) leads to the activation of heavy ion fusion reactions (such as \isotope[12]{C}+\isotope[12]{C}, \isotope[12]{C}+\isotope[16]{O}, and \isotope[16]{O}+\isotope[16]{O}) together with the \isotope[20]{Ne}+\g\ reaction, that suddenly release a large number of light particles, mostly $p$ and $\alpha$. The O burning products capture the free $p$ and $\alpha$ particles, leading to a substantial production of intermediate odd-Z elements (such as Cl, K, and Sc). For example, in these models \isotope[34]{S} is the seed for $p$ and $\alpha$ captures that produce \isotope[35]{Cl}, \isotope[39]{K}, and \isotope[45]{Sc} via \isotope[34]{S}($p$,$\gamma$)\isotope[35]{Cl}, \isotope[34]{S}($\alpha$,$\gamma$)\isotope[38]{Ar} and \isotope[38]{Ar}($\alpha$,$\gamma$)\isotope[42]{Ca}($\alpha$,$p$)\isotope[45]{Sc} or \isotope[38]{Ar}($p$,$\gamma$)\isotope[39]{K}. The enhanced number of $\alpha$ captures may also lead to the production of \isotope[44]{Ti}, that is considered to be mostly produced in explosive conditions, with a significant contribution coming from neutrino-driven explosions \citep[see, e.g.,][]{tur:10,CL17,sieverding:23,wang:24}. The production of these nuclei is independent from the metallicity, being predominantly primary, i.e., the result of the interaction of direct products of the O and C burning processes. Additionally, under these conditions, the efficient activation of the proton captures (such as those leading to the production of \isotope[35]{Cl} and \isotope[39]{K}) can significantly impact the nuclear energy generation at the base of the O burning shell. 
    Therefore, the use of an appropriately detailed nuclear network is mandatory when calculating the evolution of models experiencing C-O shell mergers. 

    At these high temperatures, the reactions start tending to an equilibrium and therefore the inverse reactions start to be efficient, in particular the photodissociation processes. As a consequence, some secondary processes (i.e., depending on pre-existing seeds in the initial composition of the star) may occur. This is the case of the production of \p-nuclei (35 neutron-deficient isotopes beyond iron, like \isotope[92]{Mo} and \isotope[130]{Ba}) via the $\gamma$--process, and weak $s$-process peak elements (such as Sr, Y, and Zr) from photodisintegrations of heavier isotopes, both previously produced by the $s$--process in He core and C shell burning or from the unburned pristine material \citep[][]{rauscher:02,ritter:18a,roberti:23,roberti:24}.

    \begin{figure}[!t]
        \includegraphics[width=0.48\textwidth]{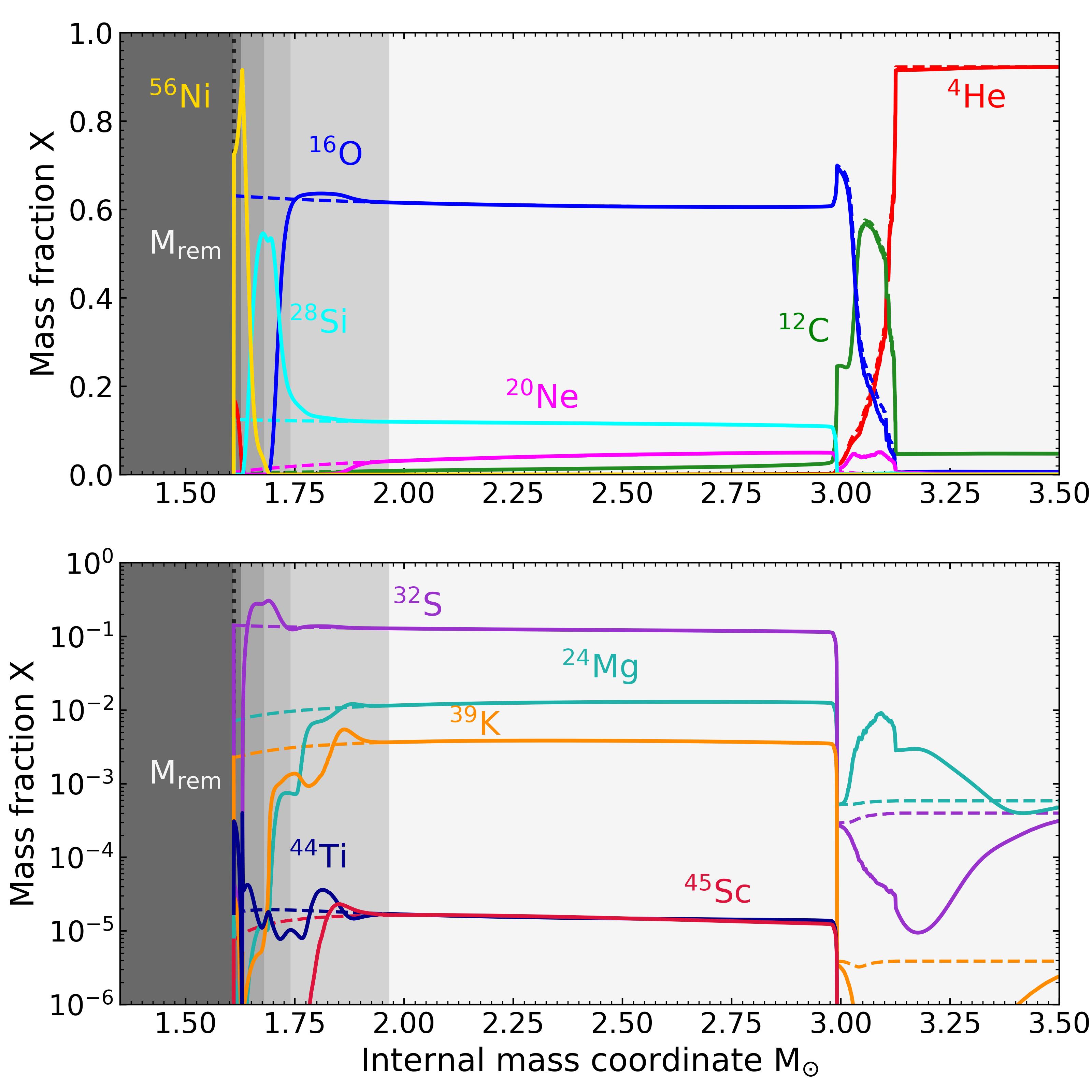}
        \caption{As \figurename~\ref{fig:x}, but for the pre-supernova (dashed lines) and post-supernova (solid lines) stages, with the inclusion of \isotope[56]{Ni}. The supernova explosion was modelled with a semi-analytical approach, using the Sedov blast wave (SBW) solution for the determination of the peak velocity of the shock throughout the stellar structure and the remnant mass prescription from \cite{fryer:12} \citep[see][for a detailed description of the explosion method]{pignatari:16,ritter:18,roberti:24b}. The gray bands in the plots mark each explosive burning stage in the corresponding mass coordinate (i.e., complete and incomplete Si burning, explosive O, Ne, and C burning). The vertical dotted line represent the mass-cut that divides the supernova ejecta from the remnant mass. The variation of \isotope[24]{Mg} and \isotope[32]{S} in the He shell represents the activation of the $\alpha$ capture chain in the explosive He burning \citep{pignatari:23}.}\label{fig:xx}
    \end{figure}  
    
    This unique nucleosynthetic signature is mostly preserved, even after the explosive nucleosynthesis, due to the extensive radius of the mixed convective zone created by the merger \citep[see \figurename~\ref{fig:xx},][and references therein]{roberti:23,roberti:24}\footnote{Note that the peak temperature of the shock scales as $\rm T_{peak}\propto R_*^{-3/4}$, with $\rm R_*$ radial coordinate of the star.}.  

\section{Stellar Archaeology} \label{sec:obs}

    As demonstrated above, the occurrence of C-O shell mergers is independent from the metallicity. Therefore we can use the observations of the earliest generation of low-mass stars to look for traces of pollution from C-O merger events in primordial massive stars. This is allowed because the interstellar cloud where most of the metal poor stars that we can still observe today were born, are thought to have experienced pollution by one or very few core-collapse supernova explosions. Therefore the observation of their surface abundances still reflect the pristine composition of the gas enriched by the first supernovae.

    \begin{figure}[!t]
        \includegraphics[width=0.48\textwidth]{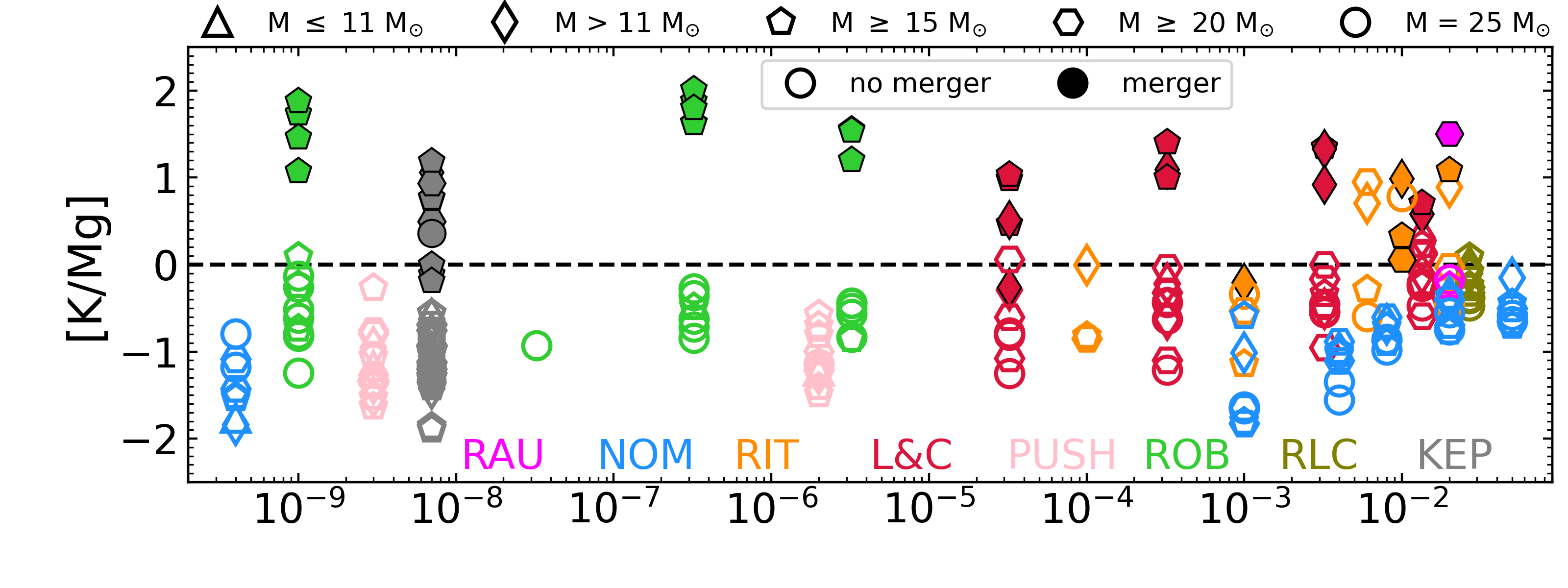}
        \includegraphics[width=0.48\textwidth]{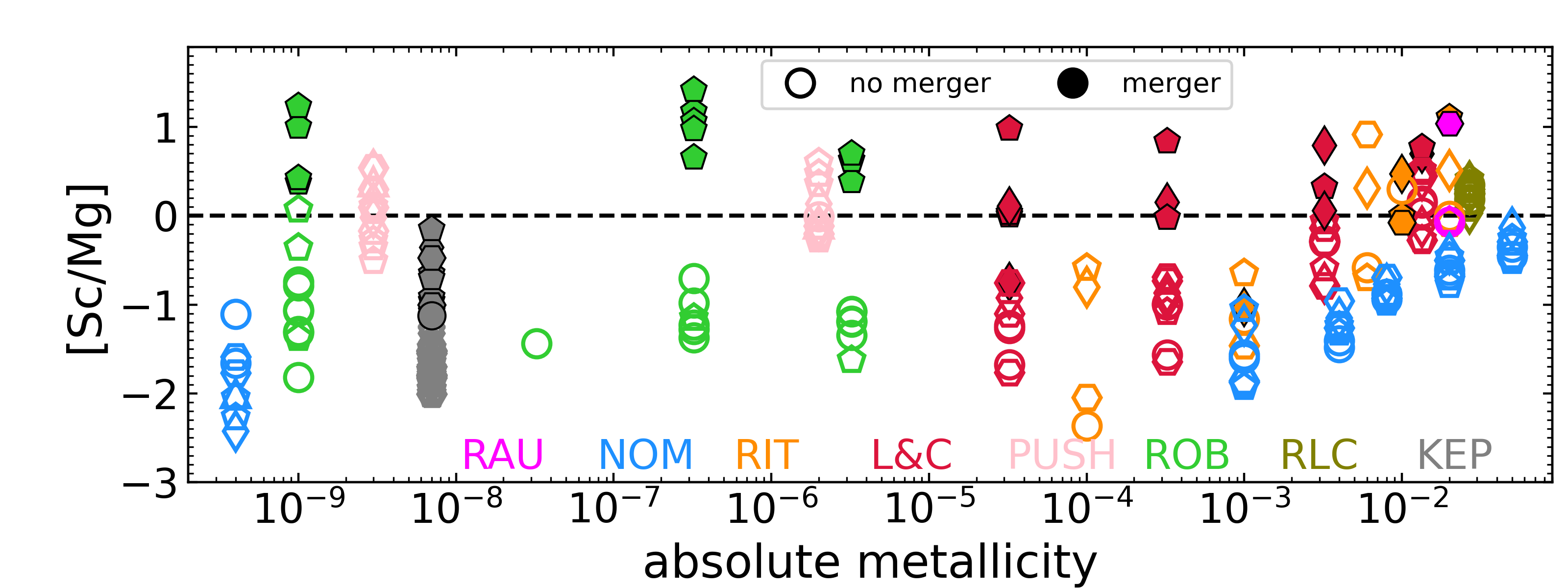}
        \caption{The [K/Mg] (upper panel) and [Sc/Mg] (lower panel) versus the initial metallicity. The adopted solar reference is from \cite{asplund:09}. The different colored labels identify each set of models (see bottom legend): RAU \cite{rauscher:02}; NOM \cite{nomoto:13}; RIT \cite{ritter:18}; L$\&$C \cite{LC18}; PUSH \cite{ebinger:20}; ROB \cite{roberti:24}; RLC \cite{RLC24}; KEP \cite{jeena:24}. Different symbols identify different initial masses (see top legend). Models with and without a C-O shell merger are represented by a filled and an empty symbol, respectively.}\label{fig:2}
    \end{figure}
    
    We explore here the use of [K/Mg] and [Sc/Mg]\footnote{[X/Y] = log$_{10}$(X/Y)-log$_{10}$(X/Y)$_{\odot}$.} as tracers for C-O mergers. As mentioned above, K and Sc are two odd-Z elements abundantly produced in a C-O shell merger, and which can be observed even at extremely low metallicity \cite{ritter:18a}. Mg, an $\alpha$ element typically produced by massive stars, can also be observed at low metallicity, but is not significantly affected by the C-O merger. Figure~\ref{fig:2} presents the [K/Mg] and [Sc/Mg] ratios as a function of the initial metallicity for most of the models shown in \figurename~\ref{fig:1}, i.e., those that provided CCSN yields. We furthermore included two other sets of models (NOM \cite{nomoto:13} and PUSH \cite{curtis:19,ebinger:20}), which, conversely, did not appear in \figurename~\ref{fig:1} because the \mco\ and \xc\ values are not available. In most of the models with C-O shell mergers, both the [K/Mg] and [Sc/Mg] are positive (i.e., their absolute ratios have from solar to super-solar values), while in most of the models without C-O shell mergers, they are negative instead (i.e., their absolute ratios have sub-solar values), especially at low metallicity, where the initial abundances of these elements are negligible. The only exceptions are the [Sc/Mg] ratios of the PUSH models, which include neutrino interactions in the explosion. The impact of neutrinos is particularly important for the $\alpha$-rich freeze out and explosive Si burning and allows these models to reach [Sc/Mg]$\gtrsim 0$ without a C-O shell merger. For further discussion of this dataset see Appendix~\ref{subapp:fig23}.

    \begin{figure*}[!t]
        \includegraphics[width=1.00\textwidth]{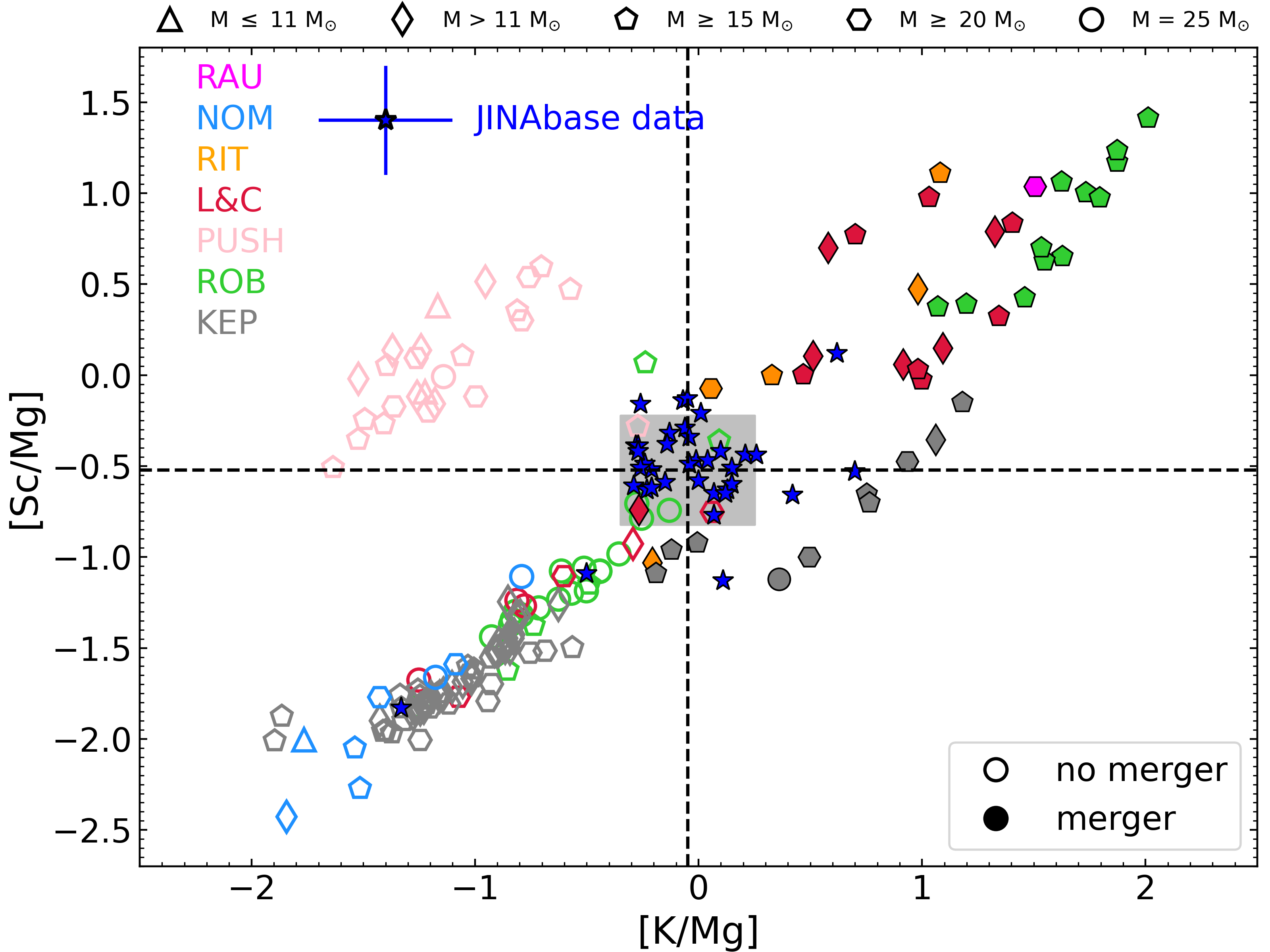}
        \caption{The [K/Mg] versus [Sc/Mg] ratios from those models selected from \figurename~\ref{fig:2} that have either a C-O shell merger or initial metallicity $\rm Z_{ini}<10^{-5}$. The adopted solar reference is from \cite{asplund:09}. The different colored labels identify each set of models as in \figurename~\ref{fig:2} (see side legend). Different symbols identify different initial masses (see top legend). Models with and without a C-O shell merger are represented by a filled and empty symbol, respectively. The blue star symbols are observations of low-mass stars of low metallicity taken from JINAbase \citep{jina} with [Fe/H]$<$--3. A representative error bar for the observation data is shown, as taken from \cite{roederer:14}. The dashed vertical and horizontal lines represent the average [K/Mg] and [Sc/Mg] ratios from the observations. The grey shaded area represents the representative error bar for the average data point.}\label{fig:3}
    \end{figure*}  
    
    Figure~\ref{fig:3} shows a comparison between the model predictions and the observations at extremely low metallicity (with [Fe/H]$\leq -3$, taken from JINAbase \citep{jina}, see the Appendix \ref{subapp:fig4}). To ensure a meaningful comparison between the observations and progenitor candidates from the sample presented in \figurename~\ref{fig:2}, we included only models with a C-O shell merger (in which the nucleosynthesis of Mg, K, and Sc is primary) and models with an initial metallicity lower than $10^{-5}$ (corresponding to $\rm [Fe/H]\approx -3$). In other words, we did not include higher metallicity models without a merger, as their ratios would be close to solar primarily due to their initial composition rather than the effects of nucleosynthesis. The observations show that the [K/Mg] ratio is centered around the solar value, whereas the [Sc/Mg] ratio is slightly sub-solar. In the plot, the grey shaded area represents the average [K/Mg] and [Sc/Mg] ratios, including an error band. This average data point divides the models into three subclasses: models without a C-O shell merger in the lower left quadrant, models with a C-O shell merger in the upper right quadrant, and models without C-O shell mergers but incorporating neutrino physics in the explosion, in the upper left quadrant. The observed data falls in a junction region between these three populations, making it difficult to identify any single population as the definitive progenitor of these second-generation stars. However, it becomes clear that a single component alone cannot reproduce the observations, suggesting that a nucleosynthetic contribution from C-O shell mergers is essential even at extremely low metallicity. 
    Currently, it is challenging to estimate the required fraction of C-O merger ejecta needed to align the interstellar cloud composition with the observed values, as this may heavily depend on the selection of progenitors without C-O shell mergers or even on the galactic chemical evolution (GCE). Nevertheless, this analysis emphasizes the need for more extensive low-metallicity observations and provides a framework for interpreting the chemical signatures of extremely low-metallicity stars based on their position within the quadrants.

\section{Discussion} \label{sec:disc}

    One of the largest sources of uncertainty in stellar models is the treatment of convection and the definition of convective borders. In this context, constraints from asteroseismology \citep[][]{aerts:03,aerts:21,burssens:23,brinkman:24} and indications from multidimensional simulations of burning shells \citep[][]{rizzuti:22,rizzuti:23,rizzuti:24} are starting to provide valuable tools to improve the future generation of 1D stellar modelling. At first glance, both seem to lead to the conclusion that mixing is more efficient at the edges of convection zones than current 1D models predict. This suggests that C-O shell mergers are a relevant feature of massive stars, and more frequent than expected.

    Another significant source of uncertainty is the determination of the reaction rates that govern the latest stages of the evolution, and in particular the \cago\, the \isotope[12]{C}+\isotope[12]{C}, and, in the case of C-O shell mergers, \isotope[12]{C}+\isotope[16]{O} and light particle capture reaction rates. The \cago\ rate is already rather well constrained by current extrapolations of available experimental data, with an estimated uncertainty of approximately $20\%$ \citep{deboer:17}. The \isotope[12]{C}+\isotope[12]{C} reaction rate is much more uncertain than the \cago\ rate and its variation can drastically change the convective history of the C burning shell and thus even affect the final fate of the star \cite[see, e.g.,][]{pignatari:13,chieffi:21}. The \isotope[12]{C}+\isotope[16]{O} reaction rate can significantly contribute to the energy generation during a C-O shell merger, especially for a low rate of C ingestion in the O shell, and, as the \isotope[12]{C}+\isotope[12]{C} reaction rate, it is still rather uncertain \citep{andrassy:20}. Future and ongoing measures of the \cago\ and the heavy ion fusion rates as in the case of the \isotope[12]{C}+\isotope[12]{C} reaction at LUNA \citep{chemseddine:24} and JUNA \citep{kajino:23}, and of the \isotope[12]{C}+\isotope[16]{O} reaction at LNS-INFN \citep{oliva:23}, will help to constrain the evolutionary properties of stars approaching the CCSN stage. 
    
    While it is estimated that most massive stars are part of binary or even multiple systems, current modeling of massive star binaries is challenging and relies on significant approximations \citep[e.g.,][]{laplace:21,brinkman:23}. Still, we can use a simplified approach to discuss C-O shell mergers in binary systems. The primary effect of binarity is a significantly larger mass loss, which can reduce the size of the helium cores and, consequently, result in smaller \mco\ and higher \xc\ compared to single stars \cite[e.g.,][]{brinkman:23}. As demonstrated in Sect. \ref{sec:c12}, a higher \xc\ would increase the entropy barriers between C, Ne, and O burning shells, thereby hindering the penetration of the convective O shell into the C- and Ne-rich zones. Therefore, we can predict that binary stars may move likely outside to the right of the grey area of \figurename~\ref{fig:1} and have less C-O shell mergers.

    The GCE effects of C-O shell mergers on odd-Z elements have been indirectly confirmed by \cite{prantzos:18}, who utilized the yield set from \cite{LC18}. In that dataset, in fact, all C-O shell mergers but one are found within rotating models (see Appendix \ref{subapp:fra}). The difference between the GCE results obtained without and with the inclusion of massive star rotating models reflects the effect of both rotation and C-O shell merger nucleosynthesis. In particular, the GCE with rotating massive stars improved the fit to observations for many elements compared to non-rotating models alone. 
    Although it is challenging to disentangle the individual contributions of rotation and C-O shell mergers to the production of odd-Z elements like K and Sc, the role of C-O mergers in their synthesis is undoubtedly as important as the effect of rotation. Notably, their GCE results for Mg, K, and Sc at extremely low metallicity align with the grey shaded area in our \figurename~\ref{fig:3}, highlighting the significant contribution of C-O mergers in matching observed values.

\section{Summary and Conclusions} \label{sec:disc}

    We studied the occurrence of C-O shell mergers in a large set of models published in literature, with a wide range of initial mass, metallicity and rotation velocity conditions. We found that, in spite of all the current uncertainties, the occurrence of C-O mergers does not depend on the initial mass, rotational velocity, nor metallicity of the star, instead, it is influenced by the intrinsic properties of the star at the end of the He burning phase, and that the representative values of \mco\ and \xc\ for models with a C-O merger are \mco $<4.90$ \msun\ and \xc $<0.277$. This provides a method to predict the occurrence of these events from the first evolutionary phases with a likelihood of about $50\%$.
    
    We furthermore showed that the nucleosynthesis during a C-O shell merger is dominated by the light particle captures, leading to a large production of the odd-Z nuclear species, such as \isotope[31]{P}, \isotope[39]{K}, \isotope[45]{Sc}, and of the radioactive species \isotope[44]{Ti}, whose Galactic origin is still unclear and debated \citep[e.g.,][and references therein]{CL17,ritter:18a,maas:22,sieverding:23}. These reactions have also a significant impact on the nuclear energy generation and they even dominate over the energy generated by the heavy ion fusions, such as \isotope[16]{O}+\isotope[16]{O} and \isotope[12]{C}+\isotope[16]{O} reactions.
    
    Finally, using observational data from stellar archaeology we have demonstrated that the impact of C-O shell mergers on nucleosynthesis was already visible in the early Universe. This highlights the significant contribution of these events to the chemical enrichment processes of galaxies and the origin of solar matter and the need to investigate them further. C-O shell mergers likely contributed to enrich the pre-Solar System material with odd-Z isotopes, such as the radioactive \isotope[40]{K}, a fundamental heat source in the evolution of rocky planets like the Earth \citep{turcotte:02,lugaro:18}.

\begin{acknowledgements}
We thank the support from the NKFI via K-project 138031 and the Lend\"ulet Program LP2023-10 of the Hungarian Academy of Sciences. LR and MP acknowledge the support to NuGrid from JINA-CEE (NSF Grant PHY-1430152) and STFC (through the University of Hull’s Consolidated Grant ST/R000840/1), and ongoing access to {\tt viper}, the University of Hull High Performance Computing Facility. LR acknowledges the support from the ChETEC-INFRA -- Transnational Access Projects 22102724-ST and 23103142-ST and the PRIN URKA Grant Number \verb |prin_2022rjlwhn|. This work was supported by the European Union’s Horizon 2020 research and innovation programme (ChETEC-INFRA -- Project no. 101008324), and the IReNA network supported by US NSF AccelNet (Grant No. OISE-1927130). ML was also supported by the NKFIH excellence grant TKP2021-NKTA-64. This work has been partially supported by the Italian grants “Premiale 2015 FIGARO” (PI: Gianluca Gemme). We acknowledge support from PRIN MUR 2022 (20224MNC5A), "Life, death and after-death of massive stars", funded by European Union - Next Generation EU.
\end{acknowledgements}

\begin{appendix}

   \onecolumn
   \section{Models used in Figure 1} \label{subapp:fig1}

    \begin{longtable}[!h]{lllcrrrr}
        \caption{List of models presented in \figurename~\ref{fig:1}. Models with a C-O shell merger have \xc\ and \mco\ in boldface.}
        \label{tab:comerger} \\
        \hline\hline
        Set & Reference & Code & $\rm M_{ini}$ (\msun) & $\rm Z_{ini}$ & $\rm V_{ini}$ ($\rm km\ s^{-1}$) & \xc & \mco\ (\msun) \\
        \hline
        \endfirsthead
        
        \multicolumn{8}{c}%
        {{\tablename\ \thetable{} -- continued from previous page}} \\
        \hline\hline
        Set & Reference & Code & $\rm M_{ini}$ (\msun) & $\rm Z_{ini}$ & $\rm V_{ini}$ ($\rm km\ s^{-1}$) & \xc\ & \mco\ (\msun) \\
        \hline
        \endhead
        
        \hline \multicolumn{8}{r}{{Continued on next page}} \\
        \endfoot
        
        \hline
        \endlastfoot
        
            ROB &\cite{roberti:24} & \verb|FRANEC| & 15.00 &  0.000e+00 &    0 & 3.62e-01 & 2.96 \\ 
            ROB &\cite{roberti:24} & \verb|FRANEC| & 15.00 &  0.000e+00 &  150 & 1.89e-01 & 4.07 \\ 
            ROB &\cite{roberti:24} & \verb|FRANEC| & 15.00 &  0.000e+00 &  300 & \textbf{1.29e-01} & \textbf{4.19} \\ 
            ROB &\cite{roberti:24} & \verb|FRANEC| & 15.00 &  0.000e+00 &  450 & \textbf{5.88e-02} & \textbf{4.67} \\ 
            ROB &\cite{roberti:24} & \verb|FRANEC| & 15.00 &  0.000e+00 &  600 & \textbf{1.37e-01} & \textbf{4.48} \\ 
            ROB &\cite{roberti:24} & \verb|FRANEC| & 15.00 &  0.000e+00 &  700 & \textbf{1.27e-01} & \textbf{4.77} \\ 
            ROB &\cite{roberti:24} & \verb|FRANEC| & 15.00 &  0.000e+00 &  800 & 1.88e-01 & 4.50 \\ 
            ROB &\cite{roberti:24} & \verb|FRANEC| & 15.00 &  3.236e-07 &    0 & 3.36e-01 & 3.08 \\ 
            ROB &\cite{roberti:24} & \verb|FRANEC| & 15.00 &  3.236e-07 &  150 & \textbf{1.09e-01} & \textbf{3.91} \\ 
            ROB &\cite{roberti:24} & \verb|FRANEC| & 15.00 &  3.236e-07 &  300 & \textbf{1.51e-01} & \textbf{3.96} \\ 
            ROB &\cite{roberti:24} & \verb|FRANEC| & 15.00 &  3.236e-07 &  450 & \textbf{1.27e-01} & \textbf{3.83} \\ 
            ROB &\cite{roberti:24} & \verb|FRANEC| & 15.00 &  3.236e-07 &  600 & \textbf{1.73e-01} & \textbf{4.31} \\ 
            ROB &\cite{roberti:24} & \verb|FRANEC| & 15.00 &  3.236e-07 &  800 & \textbf{1.21e-01} & \textbf{4.48} \\ 
            ROB &\cite{roberti:24} & \verb|FRANEC| & 15.00 &  3.236e-06 &    0 & 3.22e-01 & 3.25 \\ 
            ROB &\cite{roberti:24} & \verb|FRANEC| & 15.00 &  3.236e-06 &  300 & \textbf{1.04e-01} & \textbf{4.32} \\ 
            ROB &\cite{roberti:24} & \verb|FRANEC| & 15.00 &  3.236e-06 &  600 & \textbf{1.02e-01} & \textbf{4.16} \\ 
            ROB &\cite{roberti:24} & \verb|FRANEC| & 15.00 &  3.236e-06 &  700 & \textbf{6.23e-02} & \textbf{4.10} \\ 
            ROB &\cite{roberti:24} & \verb|FRANEC| & 25.00 &  0.000e+00 &    0 & 2.97e-01 & 5.92 \\ 
            ROB &\cite{roberti:24} & \verb|FRANEC| & 25.00 &  0.000e+00 &  150 & 1.89e-01 & 7.16 \\ 
            ROB &\cite{roberti:24} & \verb|FRANEC| & 25.00 &  0.000e+00 &  300 & 1.91e-01 & 7.52 \\ 
            ROB &\cite{roberti:24} & \verb|FRANEC| & 25.00 &  0.000e+00 &  450 & 1.78e-01 & 8.75 \\ 
            ROB &\cite{roberti:24} & \verb|FRANEC| & 25.00 &  0.000e+00 &  600 & 1.95e-01 & 9.25 \\ 
            ROB &\cite{roberti:24} & \verb|FRANEC| & 25.00 &  0.000e+00 &  700 & 7.64e-02 & 6.96 \\ 
            ROB &\cite{roberti:24} & \verb|FRANEC| & 25.00 &  0.000e+00 &  800 & 9.21e-02 & 6.89 \\ 
            ROB &\cite{roberti:24} & \verb|FRANEC| & 25.00 &  3.236e-07 &    0 & 2.78e-01 & 5.97 \\ 
            ROB &\cite{roberti:24} & \verb|FRANEC| & 25.00 &  3.236e-07 &  150 & 1.15e-01 & 6.99 \\ 
            ROB &\cite{roberti:24} & \verb|FRANEC| & 25.00 &  3.236e-07 &  300 & 1.94e-01 & 9.55 \\ 
            ROB &\cite{roberti:24} & \verb|FRANEC| & 25.00 &  3.236e-07 &  600 & 1.36e-01 & 9.58 \\ 
            ROB &\cite{roberti:24} & \verb|FRANEC| & 25.00 &  3.236e-07 &  800 & 1.78e-01 & 9.82 \\ 
            ROB &\cite{roberti:24} & \verb|FRANEC| & 25.00 &  3.236e-06 &    0 & 2.58e-01 & 7.03 \\ 
            ROB &\cite{roberti:24} & \verb|FRANEC| & 25.00 &  3.236e-06 &  300 & 1.84e-01 & 9.52 \\ 
            ROB &\cite{roberti:24} & \verb|FRANEC| & 25.00 &  3.236e-06 &  450 & 1.46e-01 & 9.69 \\ 
            ROB &\cite{roberti:24} & \verb|FRANEC| & 25.00 &  3.236e-06 &  600 & 1.56e-01 & 9.68 \\ 
            ROB &\cite{roberti:24} & \verb|FRANEC| & 25.00 &  3.236e-06 &  700 & 1.68e-01 & 9.51 \\ 
            RLC & \cite{RLC24}     & \verb|FRANEC| & 13.00 &  2.690e-02 &  300 & 2.82e-01 & 2.74 \\
            RLC & \cite{RLC24}     & \verb|FRANEC| & 13.00 &  2.690e-02 &  150 & 2.61e-01 & 2.63 \\
            RLC & \cite{RLC24}     & \verb|FRANEC| & 15.00 &  2.690e-02 &  300 & 2.70e-01 & 4.22 \\
            RLC & \cite{RLC24}     & \verb|FRANEC| & 15.00 &  2.690e-02 &  150 & 2.83e-01 & 3.19 \\
            RLC & \cite{RLC24}     & \verb|FRANEC| & 13.00 &  2.690e-02 &    0 & 3.24e-01 & 1.77 \\
            RLC & \cite{RLC24}     & \verb|FRANEC| & 15.00 &  2.690e-02 &    0 & 3.58e-01 & 2.27 \\
            RLC & \cite{RLC24}     & \verb|FRANEC| & 25.00 &  2.690e-02 &    0 & 3.02e-01 & 5.13 \\
            RLC & \cite{RLC24}     & \verb|FRANEC| & 20.00 &  2.690e-02 &    0 & 3.15e-01 & 3.61 \\
            RLC & \cite{RLC24}     & \verb|FRANEC| & 25.00 &  2.690e-02 &  300 & 2.76e-01 & 5.17 \\
            RLC & \cite{RLC24}     & \verb|FRANEC| & 25.00 &  2.690e-02 &  150 & 2.79e-01 & 5.53 \\
            RLC & \cite{RLC24}     & \verb|FRANEC| & 20.00 &  2.690e-02 &  300 & 2.72e-01 & 5.65 \\
            RLC & \cite{RLC24}     & \verb|FRANEC| & 20.00 &  2.690e-02 &  150 & 2.84e-01 & 4.47 \\
         L$\&$C &\cite{LC18}       & \verb|FRANEC| & 13.00 &  1.345e-02 &    0 & 3.72e-01 & 2.03 \\ 
         L$\&$C &\cite{LC18}       & \verb|FRANEC| & 15.00 &  1.345e-02 &    0 & 3.38e-01 & 2.78 \\ 
         L$\&$C &\cite{LC18}       & \verb|FRANEC| & 20.00 &  1.345e-02 &    0 & 3.19e-01 & 3.86 \\ 
         L$\&$C &\cite{LC18}       & \verb|FRANEC| & 25.00 &  1.345e-02 &    0 & 2.98e-01 & 6.21 \\ 
         L$\&$C &\cite{LC18}       & \verb|FRANEC| & 13.00 &  1.345e-02 &  150 & 1.87e-01 & 3.64 \\ 
         L$\&$C &\cite{LC18}       & \verb|FRANEC| & 15.00 &  1.345e-02 &  150 & 1.16e-01 & 4.79 \\ 
         L$\&$C &\cite{LC18}       & \verb|FRANEC| & 20.00 &  1.345e-02 &  150 & 2.53e-01 & 5.79 \\ 
         L$\&$C &\cite{LC18}       & \verb|FRANEC| & 25.00 &  1.345e-02 &  150 & 2.53e-01 & 6.91 \\ 
         L$\&$C &\cite{LC18}       & \verb|FRANEC| & 13.00 &  1.345e-02 &  300 & \textbf{2.06e-01} & \textbf{4.04} \\ 
         L$\&$C &\cite{LC18}       & \verb|FRANEC| & 15.00 &  1.345e-02 &  300 & \textbf{1.73e-01} & \textbf{4.59} \\ 
         L$\&$C &\cite{LC18}       & \verb|FRANEC| & 20.00 &  1.345e-02 &  300 & 2.26e-01 & 6.29 \\ %
         L$\&$C &\cite{LC18}       & \verb|FRANEC| & 25.00 &  1.345e-02 &  300 & 2.46e-01 & 7.16 \\ %
         L$\&$C &\cite{LC18}       & \verb|FRANEC| & 13.00 &  3.236e-03 &    0 & 3.37e-01 & 2.13 \\ %
         L$\&$C &\cite{LC18}       & \verb|FRANEC| & 15.00 &  3.236e-03 &    0 & 2.54e-01 & 3.21 \\ %
         L$\&$C &\cite{LC18}       & \verb|FRANEC| & 20.00 &  3.236e-03 &    0 & 3.14e-01 & 4.23 \\ %
         L$\&$C &\cite{LC18}       & \verb|FRANEC| & 25.00 &  3.236e-03 &    0 & 2.56e-01 & 6.84 \\ %
         L$\&$C &\cite{LC18}       & \verb|FRANEC| & 13.00 &  3.236e-03 &  150 & \textbf{1.82e-01} & \textbf{3.87} \\ 
         L$\&$C &\cite{LC18}       & \verb|FRANEC| & 15.00 &  3.236e-03 &  150 & 1.84e-01 & 4.76 \\ %
         L$\&$C &\cite{LC18}       & \verb|FRANEC| & 20.00 &  3.236e-03 &  150 & 1.88e-01 & 6.74 \\ %
         L$\&$C &\cite{LC18}       & \verb|FRANEC| & 25.00 &  3.236e-03 &  150 & 1.81e-01 & 8.24 \\ %
         L$\&$C &\cite{LC18}       & \verb|FRANEC| & 13.00 &  3.236e-03 &  300 & \textbf{3.79e-02} & \textbf{4.36} \\ 
         L$\&$C &\cite{LC18}       & \verb|FRANEC| & 15.00 &  3.236e-03 &  300 & \textbf{7.26e-02} & \textbf{4.89} \\ 
         L$\&$C &\cite{LC18}       & \verb|FRANEC| & 20.00 &  3.236e-03 &  300 & 1.64e-01 & 6.16 \\ 
         L$\&$C &\cite{LC18}       & \verb|FRANEC| & 25.00 &  3.236e-03 &  300 & 1.87e-01 & 9.73 \\ 
         L$\&$C &\cite{LC18}       & \verb|FRANEC| & 13.00 &  3.236e-04 &    0 & 3.35e-01 & 2.24 \\ 
         L$\&$C &\cite{LC18}       & \verb|FRANEC| & 15.00 &  3.236e-04 &    0 & 3.29e-01 & 2.73 \\ 
         L$\&$C &\cite{LC18}       & \verb|FRANEC| & 20.00 &  3.236e-04 &    0 & 3.05e-01 & 4.23 \\ 
         L$\&$C &\cite{LC18}       & \verb|FRANEC| & 25.00 &  3.236e-04 &    0 & 2.75e-01 & 5.97 \\ 
         L$\&$C &\cite{LC18}       & \verb|FRANEC| & 13.00 &  3.236e-04 &  150 & 1.87e-01 & 3.82 \\ 
         L$\&$C &\cite{LC18}       & \verb|FRANEC| & 15.00 &  3.236e-04 &  150 & \textbf{1.72e-01} & \textbf{4.08} \\ 
         L$\&$C &\cite{LC18}       & \verb|FRANEC| & 20.00 &  3.236e-04 &  150 & 1.72e-01 & 6.50 \\ %
         L$\&$C &\cite{LC18}       & \verb|FRANEC| & 25.00 &  3.236e-04 &  150 & 1.71e-01 & 8.22 \\ %
         L$\&$C &\cite{LC18}       & \verb|FRANEC| & 13.00 &  3.236e-04 &  300 & \textbf{3.07e-02} & \textbf{4.23} \\ 
         L$\&$C &\cite{LC18}       & \verb|FRANEC| & 15.00 &  3.236e-04 &  300 & \textbf{4.47e-02} & \textbf{4.72} \\ 
         L$\&$C &\cite{LC18}       & \verb|FRANEC| & 20.00 &  3.236e-04 &  300 & 5.92e-02 & 7.89 \\ %
         L$\&$C &\cite{LC18}       & \verb|FRANEC| & 25.00 &  3.236e-04 &  300 & 1.76e-01 &10.40 \\ %
         L$\&$C &\cite{LC18}       & \verb|FRANEC| & 13.00 &  3.236e-05 &    0 & 3.38e-01 & 2.25 \\ %
         L$\&$C &\cite{LC18}       & \verb|FRANEC| & 15.00 &  3.236e-05 &    0 & \textbf{2.10e-01} & \textbf{3.28} \\ 
         L$\&$C &\cite{LC18}       & \verb|FRANEC| & 20.00 &  3.236e-05 &    0 & 2.94e-01 & 4.36 \\ %
         L$\&$C &\cite{LC18}       & \verb|FRANEC| & 25.00 &  3.236e-05 &    0 & 2.48e-01 & 6.29 \\ %
         L$\&$C &\cite{LC18}       & \verb|FRANEC| & 13.00 &  3.236e-05 &  150 & \textbf{1.53e-01} & \textbf{3.85} \\ 
         L$\&$C &\cite{LC18}       & \verb|FRANEC| & 15.00 &  3.236e-05 &  150 & \textbf{1.37e-01} & \textbf{4.54} \\ 
         L$\&$C &\cite{LC18}       & \verb|FRANEC| & 20.00 &  3.236e-05 &  150 & 1.59e-01 & 5.64 \\ %
         L$\&$C &\cite{LC18}       & \verb|FRANEC| & 25.00 &  3.236e-05 &  150 & 1.73e-01 & 8.37 \\ %
         L$\&$C &\cite{LC18}       & \verb|FRANEC| & 13.00 &  3.236e-05 &  300 & \textbf{1.84e-01} & \textbf{3.82} \\ 
         L$\&$C &\cite{LC18}       & \verb|FRANEC| & 15.00 &  3.236e-05 &  300 & \textbf{1.90e-01} & \textbf{4.61} \\ 
         L$\&$C &\cite{LC18}       & \verb|FRANEC| & 20.00 &  3.236e-05 &  300 & 1.61e-01 & 6.20 \\ 
         L$\&$C &\cite{LC18}       & \verb|FRANEC| & 25.00 &  3.236e-05 &  300 & 1.82e-01 &10.50 \\
            LIM & \cite{limongi:24} & \verb|FRANEC| &2 9.22 &  1.345e-02 &    0 & 4.54e-01 & 1.08 \\
            LIM & \cite{limongi:24} & \verb|FRANEC| &5 9.25 &  1.345e-02 &    0 & 4.52e-01 & 1.10 \\
            LIM & \cite{limongi:24} & \verb|FRANEC| &0 9.30 &  1.345e-02 &    0 & 4.51e-01 & 1.11 \\
            LIM & \cite{limongi:24} & \verb|FRANEC| &0 9.50 &  1.345e-02 &    0 & 4.46e-01 & 1.16 \\
            LIM & \cite{limongi:24} & \verb|FRANEC| &0 9.80 &  1.345e-02 &    0 & 4.46e-01 & 1.20 \\
            LIM & \cite{limongi:24} & \verb|FRANEC| & 10.00 &  1.345e-02 &    0 & 4.44e-01 & 1.24 \\
            LIM & \cite{limongi:24} & \verb|FRANEC| & 11.00 &  1.345e-02 &    0 & 4.34e-01 & 1.48 \\
            LIM & \cite{limongi:24} & \verb|FRANEC| & 12.00 &  1.345e-02 &    0 & 4.13e-01 & 1.82 \\
            RIT & \cite{ritter:18}  & \verb|MESA|   & 12.00 &  1.000e-04 &  300 & 2.60e-01 & 1.55 \\ 
            RIT & \cite{ritter:18}  & \verb|MESA|   & 15.00 &  1.000e-04 &  300 & 2.34e-01 & 2.18 \\ 
            RIT & \cite{ritter:18}  & \verb|MESA|   & 20.00 &  1.000e-04 &  300 & 2.14e-01 & 3.72 \\ 
            RIT & \cite{ritter:18}  & \verb|MESA|   & 25.00 &  1.000e-04 &  300 & 2.10e-01 & 5.50 \\ 
            RIT & \cite{ritter:18}  & \verb|MESA|   & 12.00 &  1.000e-03 &  300 & \textbf{2.62e-01} & \textbf{1.49} \\ 
            RIT & \cite{ritter:18}  & \verb|MESA|   & 15.00 &  1.000e-03 &  300 & 2.34e-01 & 2.29 \\ 
            RIT & \cite{ritter:18}  & \verb|MESA|   & 20.00 &  1.000e-03 &  300 & 2.06e-01 & 3.87 \\ 
            RIT & \cite{ritter:18}  & \verb|MESA|   & 25.00 &  1.000e-03 &  300 & 2.06e-01 & 5.56 \\ 
            RIT & \cite{ritter:18}  & \verb|MESA|   & 12.00 &  6.000e-03 &  300 & 2.86e-01 & 1.42 \\ 
            RIT & \cite{ritter:18}  & \verb|MESA|   & 15.00 &  6.000e-03 &  300 & 2.61e-01 & 2.23 \\ 
            RIT & \cite{ritter:18}  & \verb|MESA|   & 20.00 &  6.000e-03 &  300 & 2.14e-01 & 3.86 \\ 
            RIT & \cite{ritter:18}  & \verb|MESA|   & 25.00 &  6.000e-03 &  300 & 2.12e-01 & 5.72 \\ 
            RIT & \cite{ritter:18}  & \verb|MESA|   & 12.00 &  1.000e-02 &  300 & \textbf{3.16e-01} & \textbf{1.29} \\ 
            RIT & \cite{ritter:18}  & \verb|MESA|   & 15.00 &  1.000e-02 &  300 & \textbf{2.77e-01} & \textbf{2.12} \\ 
            RIT & \cite{ritter:18}  & \verb|MESA|   & 20.00 &  1.000e-02 &  300 & \textbf{2.44e-01} & \textbf{3.79} \\ 
            RIT & \cite{ritter:18}  & \verb|MESA|   & 25.00 &  1.000e-02 &  300 & 2.14e-01 & 5.33 \\ 
            RIT & \cite{ritter:18}  & \verb|MESA|   & 12.00 &  2.000e-02 &  300 & 3.45e-01 & 1.18 \\ 
            RIT & \cite{ritter:18}  & \verb|MESA|   & 15.00 &  2.000e-02 &  300 & \textbf{2.67e-01} & \textbf{2.01} \\ 
            RIT & \cite{ritter:18}  & \verb|MESA|   & 20.00 &  2.000e-02 &  300 & 2.74e-01 & 3.50 \\ 
            RIT & \cite{ritter:18}  & \verb|MESA|   & 25.00 &  2.000e-02 &  300 & 2.47e-01 & 5.29 \\ 
            BRI & \cite{brinkman:19,brinkman:21} & \verb|MESA|   & 10.00 &  1.400e-02 &    0 & 3.78e-01 & 1.52 \\
            BRI & \cite{brinkman:19,brinkman:21} & \verb|MESA|   & 15.00 &  1.400e-02 &    0 & 3.23e-01 & 3.41 \\
            BRI & \cite{brinkman:19,brinkman:21} & \verb|MESA|   & 20.00 &  1.400e-02 &    0 & 2.92e-01 & 5.66 \\
            BRI & \cite{brinkman:19,brinkman:21} & \verb|MESA|   & 25.00 &  1.400e-02 &    0 & 2.62e-01 & 8.13 \\
            BRI & \cite{brinkman:19,brinkman:21} & \verb|MESA|   & 10.00 &  1.400e-02 &  150 & 3.32e-01 & 1.58 \\
            BRI & \cite{brinkman:19,brinkman:21} & \verb|MESA|   & 15.00 &  1.400e-02 &  150 & 3.11e-01 & 3.54 \\
            BRI & \cite{brinkman:19,brinkman:21} & \verb|MESA|   & 20.00 &  1.400e-02 &  150 & 2.86e-01 & 5.83 \\
            BRI & \cite{brinkman:19,brinkman:21} & \verb|MESA|   & 25.00 &  1.400e-02 &  150 & 2.71e-01 & 8.06 \\
            BRI & \cite{brinkman:19,brinkman:21} & \verb|MESA|   & 10.00 &  1.400e-02 &  300 & 3.12e-01 & 1.84 \\
            BRI & \cite{brinkman:19,brinkman:21} & \verb|MESA|   & 15.00 &  1.400e-02 &  300 & 3.13e-01 & 4.04 \\
            BRI & \cite{brinkman:19,brinkman:21} & \verb|MESA|   & 20.00 &  1.400e-02 &  300 & 2.80e-01 & 6.62 \\
            BRI & \cite{brinkman:19,brinkman:21} & \verb|MESA|   & 25.00 &  1.400e-02 &  300 & 2.63e-01 & 8.63 \\
            SIE & \cite{sieverding:18} & \verb|KEPLER| & 15.00 &  1.400e-02 &    0 & 2.25e-01 & 2.24 \\
            SIE & \cite{sieverding:18} & \verb|KEPLER| & 20.00 &  1.400e-02 &    0 & 2.15e-01 & 3.69 \\
            SIE & \cite{sieverding:18} & \verb|KEPLER| & 25.00 &  1.400e-02 &    0 & 1.89e-01 & 5.39 \\
            RAU & \cite{rauscher:02}   & \verb|KEPLER| & 15.00 &  2.000e-02 &    0 & 2.00e-01 & 2.80 \\ 
            RAU & \cite{rauscher:02}   & \verb|KEPLER| & 20.00 &  2.000e-02 &    0 & \textbf{2.03e-01} & \textbf{4.45} \\ 
            RAU & \cite{rauscher:02}   & \verb|KEPLER| & 25.00 &  2.000e-02 &    0 & 2.18e-01 & 6.44 \\ 
            KEP & \cite{jeena:24} & \verb|KEPLER| & 12.00 &  0.000e+00 &    0 & 2.99e-01 & 1.75 \\ 
            KEP & \cite{jeena:24} & \verb|KEPLER| & 12.10 &  0.000e+00 &    0 & 2.97e-01 & 1.76 \\ 
            KEP & \cite{jeena:24} & \verb|KEPLER| & 12.20 &  0.000e+00 &    0 & 2.93e-01 & 1.80 \\ 
            KEP & \cite{jeena:24} & \verb|KEPLER| & 12.30 &  0.000e+00 &    0 & 2.98e-01 & 1.85 \\ 
            KEP & \cite{jeena:24} & \verb|KEPLER| & 12.40 &  0.000e+00 &    0 & 2.96e-01 & 1.88 \\ 
            KEP & \cite{jeena:24} & \verb|KEPLER| & 12.50 &  0.000e+00 &    0 & 2.98e-01 & 1.95 \\ 
            KEP & \cite{jeena:24} & \verb|KEPLER| & 12.60 &  0.000e+00 &    0 & 2.96e-01 & 1.94 \\ 
            KEP & \cite{jeena:24} & \verb|KEPLER| & 12.70 &  0.000e+00 &    0 & 2.68e-01 & 1.96 \\ 
            KEP & \cite{jeena:24} & \verb|KEPLER| & 12.80 &  0.000e+00 &    0 & 2.60e-01 & 2.00 \\ 
            KEP & \cite{jeena:24} & \verb|KEPLER| & 12.90 &  0.000e+00 &    0 & 2.64e-01 & 2.02 \\ 
            KEP & \cite{jeena:24} & \verb|KEPLER| & 13.00 &  0.000e+00 &    0 & \textbf{2.90e-01} & \textbf{2.00} \\ 
            KEP & \cite{jeena:24} & \verb|KEPLER| & 13.10 &  0.000e+00 &    0 & 2.87e-01 & 2.05 \\ 
            KEP & \cite{jeena:24} & \verb|KEPLER| & 13.20 &  0.000e+00 &    0 & 2.87e-01 & 2.10 \\ 
            KEP & \cite{jeena:24} & \verb|KEPLER| & 13.30 &  0.000e+00 &    0 & 2.88e-01 & 2.13 \\ 
            KEP & \cite{jeena:24} & \verb|KEPLER| & 13.40 &  0.000e+00 &    0 & 2.92e-01 & 2.18 \\ 
            KEP & \cite{jeena:24} & \verb|KEPLER| & 13.50 &  0.000e+00 &    0 & 2.64e-01 & 2.20 \\ 
            KEP & \cite{jeena:24} & \verb|KEPLER| & 13.60 &  0.000e+00 &    0 & 2.82e-01 & 2.24 \\ 
            KEP & \cite{jeena:24} & \verb|KEPLER| & 13.70 &  0.000e+00 &    0 & 2.86e-01 & 2.28 \\ 
            KEP & \cite{jeena:24} & \verb|KEPLER| & 13.80 &  0.000e+00 &    0 & 2.80e-01 & 2.29 \\ 
            KEP & \cite{jeena:24} & \verb|KEPLER| & 13.90 &  0.000e+00 &    0 & 2.87e-01 & 2.30 \\ 
            KEP & \cite{jeena:24} & \verb|KEPLER| & 14.00 &  0.000e+00 &    0 & 2.85e-01 & 2.38 \\ 
            KEP & \cite{jeena:24} & \verb|KEPLER| & 14.10 &  0.000e+00 &    0 & 2.85e-01 & 2.39 \\ 
            KEP & \cite{jeena:24} & \verb|KEPLER| & 14.20 &  0.000e+00 &    0 & 2.84e-01 & 2.43 \\ 
            KEP & \cite{jeena:24} & \verb|KEPLER| & 14.30 &  0.000e+00 &    0 & 2.82e-01 & 2.59 \\ 
            KEP & \cite{jeena:24} & \verb|KEPLER| & 14.40 &  0.000e+00 &    0 & 2.84e-01 & 2.53 \\ 
            KEP & \cite{jeena:24} & \verb|KEPLER| & 14.50 &  0.000e+00 &    0 & 2.83e-01 & 2.55 \\ 
            KEP & \cite{jeena:24} & \verb|KEPLER| & 14.60 &  0.000e+00 &    0 & 2.84e-01 & 2.56 \\ 
            KEP & \cite{jeena:24} & \verb|KEPLER| & 14.70 &  0.000e+00 &    0 & 2.80e-01 & 2.59 \\ 
            KEP & \cite{jeena:24} & \verb|KEPLER| & 14.80 &  0.000e+00 &    0 & 2.78e-01 & 2.63 \\ 
            KEP & \cite{jeena:24} & \verb|KEPLER| & 14.90 &  0.000e+00 &    0 & 2.85e-01 & 2.70 \\ 
            KEP & \cite{jeena:24} & \verb|KEPLER| & 15.00 &  0.000e+00 &    0 & 2.78e-01 & 2.65 \\ 
            KEP & \cite{jeena:24} & \verb|KEPLER| & 15.20 &  0.000e+00 &    0 & 2.74e-01 & 2.68 \\ 
            KEP & \cite{jeena:24} & \verb|KEPLER| & 15.40 &  0.000e+00 &    0 & 2.83e-01 & 2.92 \\ 
            KEP & \cite{jeena:24} & \verb|KEPLER| & 15.60 &  0.000e+00 &    0 & 2.74e-01 & 2.85 \\ 
            KEP & \cite{jeena:24} & \verb|KEPLER| & 15.80 &  0.000e+00 &    0 & 2.69e-01 & 2.92 \\ 
            KEP & \cite{jeena:24} & \verb|KEPLER| & 16.00 &  0.000e+00 &    0 & 2.78e-01 & 2.98 \\ 
            KEP & \cite{jeena:24} & \verb|KEPLER| & 16.20 &  0.000e+00 &    0 & 2.65e-01 & 3.10 \\ 
            KEP & \cite{jeena:24} & \verb|KEPLER| & 16.40 &  0.000e+00 &    0 & 2.70e-01 & 3.30 \\ 
            KEP & \cite{jeena:24} & \verb|KEPLER| & 16.60 &  0.000e+00 &    0 & \textbf{2.69e-01} & \textbf{3.40} \\ 
            KEP & \cite{jeena:24} & \verb|KEPLER| & 16.80 &  0.000e+00 &    0 & 2.66e-01 & 3.25 \\ 
            KEP & \cite{jeena:24} & \verb|KEPLER| & 17.00 &  0.000e+00 &    0 & 2.66e-01 & 3.35 \\ 
            KEP & \cite{jeena:24} & \verb|KEPLER| & 17.20 &  0.000e+00 &    0 & 2.58e-01 & 3.40 \\ 
            KEP & \cite{jeena:24} & \verb|KEPLER| & 17.40 &  0.000e+00 &    0 & 2.56e-01 & 3.43 \\ 
            KEP & \cite{jeena:24} & \verb|KEPLER| & 17.60 &  0.000e+00 &    0 & 2.68e-01 & 3.64 \\ 
            KEP & \cite{jeena:24} & \verb|KEPLER| & 17.80 &  0.000e+00 &    0 & 2.59e-01 & 3.78 \\ 
            KEP & \cite{jeena:24} & \verb|KEPLER| & 18.00 &  0.000e+00 &    0 & \textbf{2.43e-01} & \textbf{3.59} \\ 
            KEP & \cite{jeena:24} & \verb|KEPLER| & 18.20 &  0.000e+00 &    0 & 2.59e-01 & 3.78 \\
            KEP & \cite{jeena:24} & \verb|KEPLER| & 18.40 &  0.000e+00 &    0 & 2.54e-01 & 3.99 \\
            KEP & \cite{jeena:24} & \verb|KEPLER| & 18.60 &  0.000e+00 &    0 & 2.49e-01 & 3.88 \\
            KEP & \cite{jeena:24} & \verb|KEPLER| & 18.80 &  0.000e+00 &    0 & \textbf{2.55e-01} & \textbf{3.88} \\ 
            KEP & \cite{jeena:24} & \verb|KEPLER| & 19.00 &  0.000e+00 &    0 & 2.51e-01 & 4.04 \\ 
            KEP & \cite{jeena:24} & \verb|KEPLER| & 19.20 &  0.000e+00 &    0 & \textbf{2.47e-01} & \textbf{4.22} \\ 
            KEP & \cite{jeena:24} & \verb|KEPLER| & 19.40 &  0.000e+00 &    0 & \textbf{2.40e-01} & \textbf{4.12} \\ 
            KEP & \cite{jeena:24} & \verb|KEPLER| & 19.60 &  0.000e+00 &    0 & \textbf{2.45e-01} & \textbf{4.35} \\ 
            KEP & \cite{jeena:24} & \verb|KEPLER| & 19.80 &  0.000e+00 &    0 & 2.44e-01 & 4.37 \\ 
            KEP & \cite{jeena:24} & \verb|KEPLER| & 20.00 &  0.000e+00 &    0 & 2.40e-01 & 4.61 \\ 
            KEP & \cite{jeena:24} & \verb|KEPLER| & 20.50 &  0.000e+00 &    0 & 2.39e-01 & 4.70 \\ 
            KEP & \cite{jeena:24} & \verb|KEPLER| & 21.00 &  0.000e+00 &    0 & \textbf{2.37e-01} & \textbf{4.90} \\ 
            KEP & \cite{jeena:24} & \verb|KEPLER| & 21.50 &  0.000e+00 &    0 & 2.32e-01 & 5.04 \\ 
            KEP & \cite{jeena:24} & \verb|KEPLER| & 22.00 &  0.000e+00 &    0 & 2.28e-01 & 5.59 \\ 
            KEP & \cite{jeena:24} & \verb|KEPLER| & 22.50 &  0.000e+00 &    0 & 2.28e-01 & 5.38 \\ 
            KEP & \cite{jeena:24} & \verb|KEPLER| & 23.00 &  0.000e+00 &    0 & 2.25e-01 & 5.74 \\ 
            KEP & \cite{jeena:24} & \verb|KEPLER| & 23.50 &  0.000e+00 &    0 & 2.24e-01 & 6.11 \\ 
            KEP & \cite{jeena:24} & \verb|KEPLER| & 24.00 &  0.000e+00 &    0 & 2.21e-01 & 6.36 \\ 
            KEP & \cite{jeena:24} & \verb|KEPLER| & 24.50 &  0.000e+00 &    0 & \textbf{2.18e-01} & \textbf{6.49} \\ 
            KEP & \cite{jeena:24} & \verb|KEPLER| & 25.00 &  0.000e+00 &    0 & \textbf{2.14e-01} & \textbf{6.53} \\ 
            PGN & \cite{pignatari:16} & \verb|GENEC|  & 15.00 &  2.000e-02 &    0 & 3.54e-01 & 2.78 \\
            PGN & \cite{pignatari:16} & \verb|GENEC|  & 20.00 &  2.000e-02 &    0 & 3.05e-01 & 4.49 \\
            PGN & \cite{pignatari:16} & \verb|GENEC|  & 25.00 &  2.000e-02 &    0 & 2.69e-01 & 6.12 \\
            \hline
    \end{longtable}
    \twocolumn
    
        In the following, we list the prescriptions used for convection and for the \cago\ rate in the 209 models shown in \figurename~\ref{fig:1}. The adopted \isotope[12]{C}+\isotope[12]{C} rate is instead the same for all the models and it is the one provided by \cite{caughlan:88}. We refer the reader to the cited papers for a more detailed description of each set of models. The models with a C-O shell merger are listed in \tablename~\ref{tab:comerger}.

        \subsection{The KEPLER models (RAU, SIE, KEP)}
        
            We considered models calculated with the \verb|KEPLER| code  from \cite{rauscher:02} (RAU) and \cite{sieverding:18} (SIE), with initial mass equal to 15, 20, and 25 \msun\ and solar metallicity \citep[][with Z = 0.02]{anders:89}. Convective borders are defined by the Ledoux criterion, as in \cite{weaver:78}. The adopted \cago\ rate is from \cite{buchmann:96}, multiplied by a factor of 1.2. In addition to RAU and SIE sets, we also consider the \verb|KEPLER| models by \cite{jeena:24} (KEP). The KEP set includes 76 models between 12 and 30 \msun\ at zero metallicity, with the primordial Big Bang nucleosynthesis (BBN) initial composition from \cite{cyburt:02}. For consistency with most of the other sets having initial masses up to 25 \msun, we only consider the 66 KEP models with an initial mass $\rm M_{ini} \leq 25$ \msun. The convection and mixing criteria are the same as in \cite{heger:10} and consist in Ledoux with a semiconvective diffusion coefficient that is roughly 10$\%$ of the radiative diffusion coefficient, plus a small amount of convective overshooting included by forcing convective boundary zones to be semiconvective. The \cago\ rate is the same as adopted by RAU and SIE.

            \begin{figure}[!t]
                \includegraphics[width=0.48\textwidth]{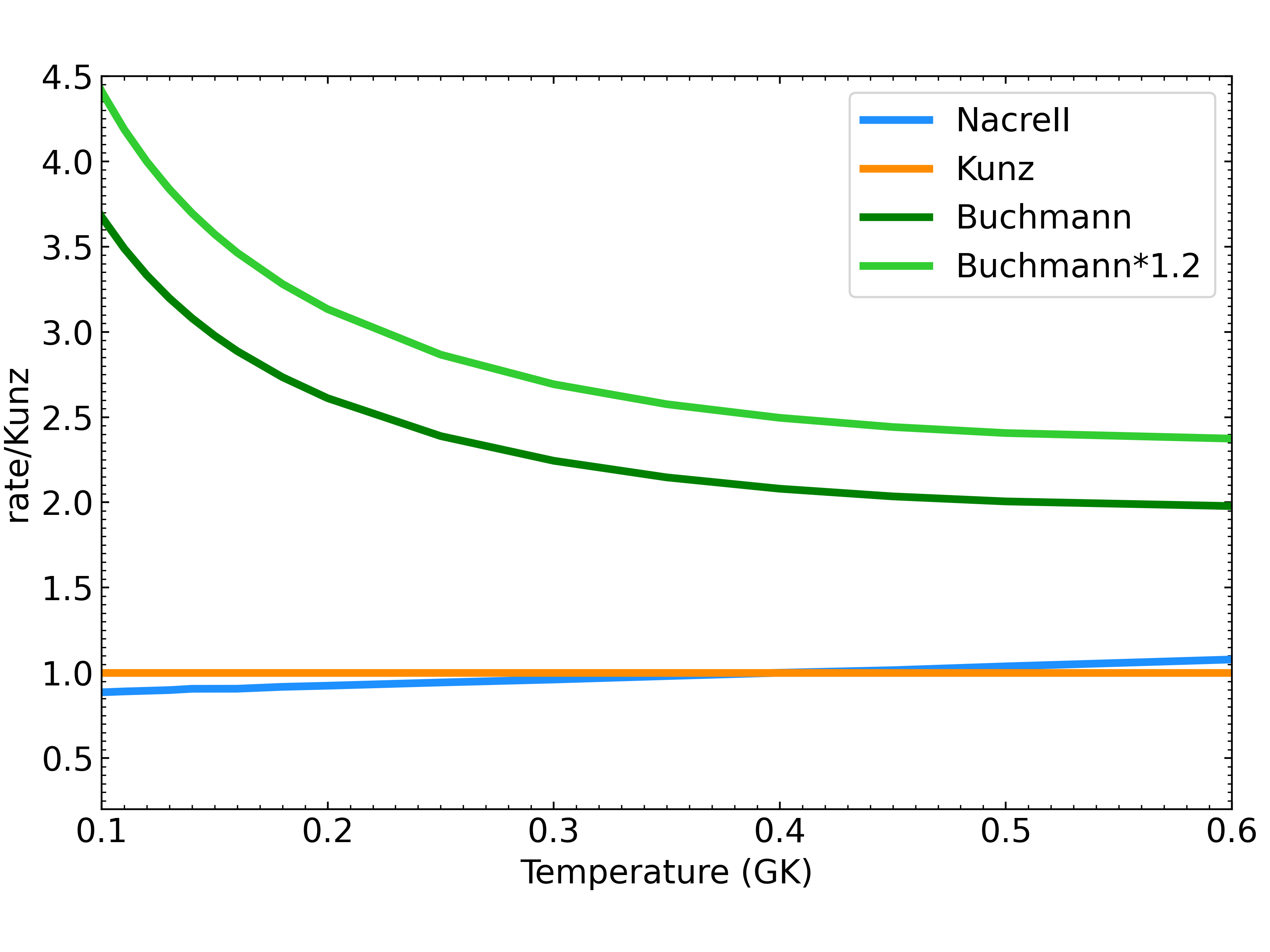}
                \caption{The different \cago\ reaction rates normalized to the \cite{kunz:02} reaction rate in the typical He burning temperature range.}\label{fig:4}
            \end{figure} 
        
            We note that the \cago\ rate of \cite{buchmann:96} is a factor of 2-3 higher than that of \cite{kunz:02}, adopted by most of the other sets of models (see below), in the typical He-burning temperature range ($\rm T_9 = 0.2-0.5$, where $\rm T_9 = T/10^9\ K$, \figurename~\ref{fig:4}). As a consequence, the \mco\ versus \xc\ datapoints from the for \verb|KEPLER| models are shifted toward lower \xc\ values than the other models in \figurename~\ref{fig:1}.

        \subsection{GENEC models (PGN)}

            We include the 15, 20, and 25 \msun\ models with solar metallicity \citep[Z=0.02,][]{grevesse:93} presented in \cite{pignatari:16} (PGN), computed using the \verb|GENEC| stellar evolutionary code \cite{eggenberger:08}. The evolution of these models is calculated up to the end of central Si burning phase, while the birth of the Si burning shell and the pre-collapse phase are not included. For this reason, we do not know if these models may experience a C-O shell merger in their final stages, while the structure of these models at the end of the calculated evolution suggest that a C-O shell merger is very unlikely, as the external border of the O-burning shell is rather distant in mass from the base of the C-burning shell. The adopted convection criterion is Schwarzschild, with core overshooting included only in the central H and He burning phases (with an overshooting parameter $\rm \alpha_{ov} = 0.2 H_p$). The \cago\ rate is from \cite{kunz:02}.

        \subsection{FRANEC models (L$\&$C, RLC, LIM, ROB)} \label{subapp:fra}
    
            We include several sets of models calculated with the \verb|FRANEC| code \citep[][]{CL13,LC18}. The L$\&$C \citep{LC18} and RLC \citep{RLC24} models are part of a large set of rotating massive stars, with initial masses between 13 and 120 \msun\ and spanning 5 different initial metallicities, $\rm [Fe/H]=0.3, 0,-1,-2,-3$. In this work, we considered only the models up to $\rm M_{ini} \leq 25$ \msun\ (i.e., $\rm M_{ini}$ = 13, 15, 20, and 25 \msun), at the three available rotation velocities (0, 150, and 300 km/s), and at all the metallicities, for a total of 60 models, 12 of which present C-O shell mergers. In these sets, the borders of the convective zones are defined according to the Ledoux criterion. In addition, $\rm 0.2 H_p$ of overshooting is included at the outer edge of the convective core only during the core H-burning phase. The \cago\ rate is taken from \cite{kunz:02}. 
            The LIM set \citep{limongi:24} includes models at solar metallicity in the range between 7 and 12 \msun. Since we are interested in the CCSN progenitor evolution, we consider only the models that evolve to the formation of a Fe core, corresponding to models from $\rm M_{ini}$ = 9.22 \msun. In this set, the convective borders are determined according to the Ledoux criterion in the H-rich layers and according to the Schwarzschild criterion elsewhere. Core overshooting is included during core H burning with $\rm \alpha_{ov} = 0.2 H_p$. The adopted \cago\ rate is from \cite{nacreii}, a re-analysis that also includes the experimental points from \cite{kunz:02}. In typical He burning conditions it is very similar to the \cite{kunz:02} rate (\figurename~\ref{fig:4}). None of these models undergo C-O shell mergers. The same prescriptions for convection and \cago\ reaction rate are also adopted in the ROB set \cite{roberti:24}. This set includes 34 models of 15 and 25 \msun\ stars at zero and extremely low metallicity ($\rm [Fe/H] \leq -4$) spanning a wide range of initial rotation velocity (from 0 to 800 $\rm km\ s^{-1}$). All but two of the rotating 15 \msun\ models undergo C-O shell mergers.

        \subsection{MESA models (RIT, BRI)}

            We include two sets of models calculated with the \verb|MESA| code \cite{paxton:18}. The RIT set \cite{ritter:18} includes massive star models of 12, 15, 20, 25 \msun, at 5 different metallicities: Z = 0.02, 0.01, 0.006, 0.001, 0.0001. The exponential diffusion model of \cite{freytag:96} defines all the external convective boundaries from the pre-main sequence up to the end of core He burning. The reaction rate database is the same as used in PGN models. Five RIT models experience C-O shell mergers. The BRI set \citep{brinkman:19,brinkman:21} includes non-rotating and rotating massive star models at solar metallicity with $\rm M_{ini}$ between 10 and 80 \msun\ and $\rm v_{ini}$ = 0, 150, and 300 $\rm km\ s^{-1}$, calculated up to the collapse of the Fe core. Again, we only considered models up to 25 \msun\ (i.e., 10, 15, 20, and 25 \msun). The convective boundaries are determined using the Ledoux criterion, with overshooting included only for the central burning stages ($\rm \alpha_{ov} = 0.2 H_p$) and in the H burning shell ( $\rm \alpha_{ov} = 0.1 H_p$). The adopted \cago\ rate is \cite{kunz:02}. None of these models experience C-O shell mergers.

    \onecolumn
    \section{Models used in Figures 4 and 5} \label{subapp:fig23}
   
    \begin{longtable}{llcrrrr}
        \caption{List of models presented in \figurename~\ref{fig:3}. Models with a C-O shell merger have [K/Mg] and [Sc/Mg] in boldface.}
        \label{tab:kscmg} \\
        \hline\hline
        Set & Reference & $\rm M_{ini}$ (\msun) & $\rm Z_{ini}$ & $\rm V_{ini}$ ($\rm km\ s^{-1}$) & [K/Mg] & [Sc/Mg] \\
        \hline
        \endfirsthead
        
        \multicolumn{7}{c}
        {{\tablename\ \thetable{} -- continued from previous page}} \\
        \hline\hline
        Set & Reference & $\rm M_{ini}$ (\msun) & $\rm Z_{ini}$ & $\rm V_{ini}$ ($\rm km\ s^{-1}$) & [K/Mg] & [Sc/Mg] \\
        \hline
        \endhead
        
        \hline \multicolumn{7}{r}{{Continued on next page}} \\
        \endfoot
        
        \hline
        \endlastfoot

         ROB & \cite{roberti:24} & 25.00 & 3.236e-08 &    0 & -0.93 & -1.44 \\ 
         ROB & \cite{roberti:24} & 25.00 & 3.236e-07 &    0 & -0.71 & -1.28 \\ 
         ROB & \cite{roberti:24} & 15.00 & 3.236e-06 &    0 & -0.85 & -1.62 \\ 
         ROB & \cite{roberti:24} & 25.00 & 0.000e+00 &    0 & -1.25 & -1.82 \\ 
         ROB & \cite{roberti:24} & 15.00 & 0.000e+00 &  800 & -0.24 &  0.07 \\ 
         ROB & \cite{roberti:24} & 15.00 & 3.236e-07 &  600 & \textbf{ 1.63} &  \textbf{0.65} \\ 
         ROB & \cite{roberti:24} & 15.00 & 3.236e-07 &  450 &  \textbf{1.87} & \textbf{1.17} \\ 
         ROB & \cite{roberti:24} & 15.00 & 0.000e+00 &  450 &  \textbf{1.07} &  \textbf{0.37} \\ 
         ROB & \cite{roberti:24} & 25.00 & 3.236e-06 &  600 & -0.57 & -1.20 \\ 
         ROB & \cite{roberti:24} & 15.00 & 0.000e+00 &  600 &  \textbf{1.73} &  \textbf{1.00} \\ 
         ROB & \cite{roberti:24} & 25.00 & 3.236e-06 &  450 & -0.44 & -1.08 \\ 
         ROB & \cite{roberti:24} & 15.00 & 3.236e-07 &  800 &  \textbf{1.62} &  \textbf{1.06} \\ 
         ROB & \cite{roberti:24} & 15.00 & 3.236e-06 &  700 &  \textbf{1.20} &  \textbf{0.39} \\ 
         ROB & \cite{roberti:24} & 25.00 & 3.236e-07 &  150 & -0.28 & -0.71 \\ 
         ROB & \cite{roberti:24} & 25.00 & 3.236e-07 &  300 & -0.84 & -1.37 \\ 
         ROB & \cite{roberti:24} & 25.00 & 0.000e+00 &  700 & -0.13 & -0.74 \\ 
         ROB & \cite{roberti:24} & 25.00 & 0.000e+00 &  300 & -0.51 & -1.06 \\ 
         ROB & \cite{roberti:24} & 25.00 & 0.000e+00 &  150 & -0.61 & -1.08 \\ 
         ROB & \cite{roberti:24} & 15.00 & 3.236e-06 &  300 &  \textbf{1.55} &  \textbf{0.63} \\ 
         ROB & \cite{roberti:24} & 15.00 & 0.000e+00 &  300 &  \textbf{1.87} &  \textbf{1.23} \\ 
         ROB & \cite{roberti:24} & 15.00 & 0.000e+00 &  150 &  0.09 & -0.36 \\ 
         ROB & \cite{roberti:24} & 25.00 & 3.236e-06 &  300 & -0.84 & -1.34 \\ 
         ROB & \cite{roberti:24} & 25.00 & 3.236e-06 &  700 & -0.44 & -1.08 \\ 
         ROB & \cite{roberti:24} & 15.00 & 3.236e-07 &  150 &  \textbf{2.01} &  \textbf{1.41} \\ 
         ROB & \cite{roberti:24} & 15.00 & 3.236e-07 &  300 &  \textbf{1.80} &  \textbf{0.97} \\ 
         ROB & \cite{roberti:24} & 15.00 & 0.000e+00 &  700 &  \textbf{1.46} &  \textbf{0.42} \\ 
         ROB & \cite{roberti:24} & 25.00 & 0.000e+00 &  450 & -0.82 & -1.30 \\ 
         ROB & \cite{roberti:24} & 15.00 & 3.236e-06 &  600 &  \textbf{1.53} &  \textbf{0.70} \\ 
         ROB & \cite{roberti:24} & 25.00 & 0.000e+00 &  600 & -0.79 & -1.32 \\ 
         ROB & \cite{roberti:24} & 25.00 & 3.236e-07 &  800 & -0.63 & -1.23 \\ 
         ROB & \cite{roberti:24} & 25.00 & 0.000e+00 &  800 & -0.26 & -0.79 \\ 
         ROB & \cite{roberti:24} & 25.00 & 3.236e-07 &  600 & -0.36 & -0.98 \\ 
         ROB & \cite{roberti:24} & 25.00 & 3.236e-06 &    0 & -0.50 & -1.18 \\ 
         ROB & \cite{roberti:24} & 15.00 & 0.000e+00 &    0 & -0.74 & -1.38 \\ 
         ROB & \cite{roberti:24} & 15.00 & 3.236e-07 &    0 & -0.49 & -1.15 \\ 
         L$\&$C  & \cite{LC18}    & 13.00 & 1.345e-02 &  300 &  \textbf{0.58} &  \textbf{0.70} \\ 
         L$\&$C  & \cite{LC18}    & 15.00 & 1.345e-02 &  300 &  \textbf{0.70} & \textbf{ 0.77} \\ 
         L$\&$C  & \cite{LC18}    & 15.00 & 3.236e-03 &  300 &  \textbf{1.34} & \textbf{ 0.32} \\ 
         L$\&$C  & \cite{LC18}    & 13.00 & 3.236e-03 &  150 &  \textbf{1.33} & \textbf{ 0.79} \\ 
         L$\&$C  & \cite{LC18}    & 13.00 & 3.236e-03 &  300 &  \textbf{0.92} & \textbf{ 0.06} \\ 
         L$\&$C  & \cite{LC18}    & 13.00 & 3.236e-04 &  300 &  \textbf{1.09} & \textbf{ 0.15} \\ 
         L$\&$C  & \cite{LC18}    & 15.00 & 3.236e-04 &  150 &  \textbf{1.40} & \textbf{ 0.83} \\ 
         L$\&$C  & \cite{LC18}    & 15.00 & 3.236e-04 &  300 &  \textbf{1.00} & \textbf{-0.03} \\ 
         L$\&$C  & \cite{LC18}    & 15.00 & 3.236e-05 &    0 &  \textbf{0.47} & \textbf{ 0.00} \\ 
         L$\&$C  & \cite{LC18}    & 13.00 & 3.236e-05 &    0 & -0.29 & -0.93 \\ 
         L$\&$C  & \cite{LC18}    & 15.00 & 3.236e-05 &  150 & \textbf{ 0.98} & \textbf{ 0.03} \\ 
         L$\&$C  & \cite{LC18}    & 15.00 & 3.236e-05 &  300 & \textbf{ 1.03} & \textbf{ 0.98} \\ 
         L$\&$C  & \cite{LC18}    & 13.00 & 3.236e-05 &  150 & \textbf{-0.27} & \textbf{-0.74} \\ 
         L$\&$C  & \cite{LC18}    & 13.00 & 3.236e-05 &  300 & \textbf{ 0.51} & \textbf{ 0.10} \\ 
         L$\&$C  & \cite{LC18}    & 20.00 & 3.236e-05 &  150 & -0.60 & -1.10 \\ 
         L$\&$C  & \cite{LC18}    & 20.00 & 3.236e-05 &  300 &  0.06 & -0.75 \\ 
         L$\&$C  & \cite{LC18}    & 25.00 & 3.236e-05 &  150 & -0.81 & -1.24 \\ 
         L$\&$C  & \cite{LC18}    & 25.00 & 3.236e-05 &  300 & -0.78 & -1.27 \\ 
         L$\&$C  & \cite{LC18}    & 20.00 & 3.236e-05 &    0 & -1.07 & -1.77 \\ 
         L$\&$C  & \cite{LC18}    & 25.00 & 3.236e-05 &    0 & -1.25 & -1.68 \\ 
         RIT & \cite{ritter:18}  & 15.00 & 2.000e-02 &    0 & \textbf{ 1.08} & \textbf{ 1.11} \\ 
         RIT & \cite{ritter:18}  & 12.00 & 1.000e-02 &    0 & \textbf{ 0.98} & \textbf{ 0.47} \\ 
         RIT & \cite{ritter:18}  & 15.00 & 1.000e-02 &    0 & \textbf{ 0.33} & \textbf{-0.00} \\ 
         RIT & \cite{ritter:18}  & 20.00 & 1.000e-02 &    0 & \textbf{ 0.06} & \textbf{-0.07} \\ 
         RIT & \cite{ritter:18}  & 12.00 & 1.000e-03 &    0 & \textbf{-0.21} & \textbf{-1.03} \\ 
         RAU & \cite{rauscher:02} & 20.00 & 2.000e-02 &    0 & \textbf{ 1.51} & \textbf{ 1.04} \\ 
         KEP & \cite{jeena:24} & 12.00 & 0.000e+00 &    0 & -0.86 & -1.45 \\ 
         KEP & \cite{jeena:24} & 12.10 & 0.000e+00 &    0 & -0.89 & -1.48 \\ 
         KEP & \cite{jeena:24} & 12.20 & 0.000e+00 &    0 & -0.91 & -1.51 \\ 
         KEP & \cite{jeena:24} & 12.30 & 0.000e+00 &    0 & -0.87 & -1.50 \\ 
         KEP & \cite{jeena:24} & 12.40 & 0.000e+00 &    0 & -0.86 & -1.49 \\ 
         KEP & \cite{jeena:24} & 12.50 & 0.000e+00 &    0 & -0.90 & -1.52 \\ 
         KEP & \cite{jeena:24} & 12.60 & 0.000e+00 &    0 & -0.85 & -1.47 \\ 
         KEP & \cite{jeena:24} & 12.70 & 0.000e+00 &    0 & -0.85 & -1.25 \\ 
         KEP & \cite{jeena:24} & 12.80 & 0.000e+00 &    0 & -0.81 & -1.33 \\ 
         KEP & \cite{jeena:24} & 12.90 & 0.000e+00 &    0 & -0.80 & -1.32 \\ 
         KEP & \cite{jeena:24} & 13.00 & 0.000e+00 &    0 &  \textbf{1.06} & \textbf{-0.36} \\ 
         KEP & \cite{jeena:24} & 13.10 & 0.000e+00 &    0 & -0.91 & -1.52 \\ 
         KEP & \cite{jeena:24} & 13.20 & 0.000e+00 &    0 & -0.63 & -1.26 \\ 
         KEP & \cite{jeena:24} & 13.30 & 0.000e+00 &    0 & -0.84 & -1.50 \\ 
         KEP & \cite{jeena:24} & 13.40 & 0.000e+00 &    0 & -0.83 & -1.45 \\ 
         KEP & \cite{jeena:24} & 13.50 & 0.000e+00 &    0 & -0.82 & -1.43 \\ 
         KEP & \cite{jeena:24} & 13.60 & 0.000e+00 &    0 & -1.01 & -1.64 \\ 
         KEP & \cite{jeena:24} & 13.70 & 0.000e+00 &    0 & -1.06 & -1.69 \\ 
         KEP & \cite{jeena:24} & 13.80 & 0.000e+00 &    0 & -1.01 & -1.67 \\ 
         KEP & \cite{jeena:24} & 13.90 & 0.000e+00 &    0 & -1.14 & -1.75 \\ 
         KEP & \cite{jeena:24} & 14.00 & 0.000e+00 &    0 & -1.16 & -1.77 \\ 
         KEP & \cite{jeena:24} & 14.10 & 0.000e+00 &    0 & -1.10 & -1.72 \\ 
         KEP & \cite{jeena:24} & 14.20 & 0.000e+00 &    0 & -1.14 & -1.75 \\ 
         KEP & \cite{jeena:24} & 14.30 & 0.000e+00 &    0 & -1.23 & -1.83 \\ 
         KEP & \cite{jeena:24} & 14.40 & 0.000e+00 &    0 & -1.20 & -1.78 \\ 
         KEP & \cite{jeena:24} & 14.50 & 0.000e+00 &    0 & -1.27 & -1.83 \\ 
         KEP & \cite{jeena:24} & 14.60 & 0.000e+00 &    0 & -1.25 & -1.84 \\ 
         KEP & \cite{jeena:24} & 14.70 & 0.000e+00 &    0 & -1.20 & -1.77 \\ 
         KEP & \cite{jeena:24} & 14.80 & 0.000e+00 &    0 & -1.43 & -1.90 \\ 
         KEP & \cite{jeena:24} & 14.90 & 0.000e+00 &    0 & -1.28 & -1.87 \\ 
         KEP & \cite{jeena:24} & 15.00 & 0.000e+00 &    0 & -1.23 & -1.81 \\ 
         KEP & \cite{jeena:24} & 15.20 & 0.000e+00 &    0 & -1.25 & -1.73 \\ 
         KEP & \cite{jeena:24} & 15.40 & 0.000e+00 &    0 & -1.37 & -1.97 \\ 
         KEP & \cite{jeena:24} & 15.60 & 0.000e+00 &    0 & -1.32 & -1.90 \\ 
         KEP & \cite{jeena:24} & 15.80 & 0.000e+00 &    0 & -1.22 & -1.81 \\ 
         KEP & \cite{jeena:24} & 16.00 & 0.000e+00 &    0 & -1.40 & -1.96 \\ 
         KEP & \cite{jeena:24} & 16.20 & 0.000e+00 &    0 & -1.25 & -1.84 \\ 
         KEP & \cite{jeena:24} & 16.40 & 0.000e+00 &    0 & -1.41 & -1.96 \\ 
         KEP & \cite{jeena:24} & 16.60 & 0.000e+00 &    0 &  \textbf{1.18} & \textbf{-0.15} \\ 
         KEP & \cite{jeena:24} & 16.80 & 0.000e+00 &    0 & -1.21 & -1.78 \\ 
         KEP & \cite{jeena:24} & 17.00 & 0.000e+00 &    0 & -1.86 & -1.88 \\ 
         KEP & \cite{jeena:24} & 17.20 & 0.000e+00 &    0 & -1.33 & -1.76 \\ 
         KEP & \cite{jeena:24} & 17.40 & 0.000e+00 &    0 & -1.25 & -1.76 \\ 
         KEP & \cite{jeena:24} & 17.60 & 0.000e+00 &    0 & -1.90 & -2.01 \\ 
         KEP & \cite{jeena:24} & 17.80 & 0.000e+00 &    0 & -1.16 & -1.75 \\ 
         KEP & \cite{jeena:24} & 18.00 & 0.000e+00 &    0 & \textbf{-0.12} & \textbf{-0.96} \\ 
         KEP & \cite{jeena:24} & 18.20 & 0.000e+00 &    0 & -1.33 & -1.83 \\ 
         KEP & \cite{jeena:24} & 18.40 & 0.000e+00 &    0 & -1.03 & -1.60 \\ 
         KEP & \cite{jeena:24} & 18.60 & 0.000e+00 &    0 & -1.00 & -1.62 \\ 
         KEP & \cite{jeena:24} & 18.80 & 0.000e+00 &    0 & \textbf{-0.01} & \textbf{-0.92} \\ 
         KEP & \cite{jeena:24} & 19.00 & 0.000e+00 &    0 & -1.01 & -1.66 \\ 
         KEP & \cite{jeena:24} & 19.20 & 0.000e+00 &    0 & \textbf{-0.19} & \textbf{-1.09} \\ 
         KEP & \cite{jeena:24} & 19.40 & 0.000e+00 &    0 & \textbf{ 0.75} & \textbf{-0.65} \\ 
         KEP & \cite{jeena:24} & 19.60 & 0.000e+00 &    0 & \textbf{ 0.77} & \textbf{-0.70} \\ 
         KEP & \cite{jeena:24} & 19.80 & 0.000e+00 &    0 & -0.56 & -1.50 \\ 
         KEP & \cite{jeena:24} & 20.00 & 0.000e+00 &    0 & -0.93 & -1.55 \\ 
         KEP & \cite{jeena:24} & 20.50 & 0.000e+00 &    0 & -0.69 & -1.51 \\ 
         KEP & \cite{jeena:24} & 21.00 & 0.000e+00 &    0 &  \textbf{0.93} & \textbf{-0.47} \\ 
         KEP & \cite{jeena:24} & 21.50 & 0.000e+00 &    0 & -1.20 & -1.83 \\ 
         KEP & \cite{jeena:24} & 22.00 & 0.000e+00 &    0 & -0.75 & -1.52 \\ 
         KEP & \cite{jeena:24} & 22.50 & 0.000e+00 &    0 & -0.93 & -1.70 \\ 
         KEP & \cite{jeena:24} & 23.00 & 0.000e+00 &    0 & -1.12 & -1.80 \\ 
         KEP & \cite{jeena:24} & 23.50 & 0.000e+00 &    0 & -1.25 & -2.00 \\ 
         KEP & \cite{jeena:24} & 24.00 & 0.000e+00 &    0 & -0.94 & -1.79 \\ 
         KEP & \cite{jeena:24} & 24.50 & 0.000e+00 &    0 &  \textbf{0.50} & \textbf{-1.00} \\ 
         KEP & \cite{jeena:24} & 25.00 & 0.000e+00 &    0 &  \textbf{0.36} & \textbf{-1.12} \\ 
        PUSH &\cite{ebinger:20} & 11.00 & 2.000e-06 &    0 & -1.22 & -0.09 \\ 
        PUSH &\cite{ebinger:20} & 12.00 & 2.000e-06 &    0 & -1.26 & -0.12 \\ 
        PUSH &\cite{ebinger:20} & 13.00 & 2.000e-06 &    0 & -1.24 &  0.14 \\ 
        PUSH &\cite{ebinger:20} & 14.00 & 2.000e-06 &    0 & -1.18 & -0.16 \\ 
        PUSH &\cite{ebinger:20} & 15.00 & 2.000e-06 &    0 & -1.49 & -0.25 \\ 
        PUSH &\cite{ebinger:20} & 16.00 & 2.000e-06 &    0 & -1.41 & -0.27 \\ 
        PUSH &\cite{ebinger:20} & 17.00 & 2.000e-06 &    0 & -0.70 &  0.60 \\ 
        PUSH &\cite{ebinger:20} & 18.00 & 2.000e-06 &    0 & -0.57 &  0.47 \\ 
        PUSH &\cite{ebinger:20} & 19.00 & 2.000e-06 &    0 & -0.81 &  0.35 \\ 
        PUSH &\cite{ebinger:20} & 20.00 & 2.000e-06 &    0 & -1.21 & -0.20 \\ 
        PUSH &\cite{ebinger:20} & 24.00 & 2.000e-06 &    0 & -1.00 & -0.12 \\ 
        PUSH &\cite{ebinger:20} & 25.00 & 2.000e-06 &    0 & -1.14 & -0.01 \\ 
        PUSH &\cite{ebinger:20} & 11.00 & 0.000e+00 &    0 & -1.17 &  0.38 \\ 
        PUSH &\cite{ebinger:20} & 12.00 & 0.000e+00 &    0 & -0.95 &  0.51 \\ 
        PUSH &\cite{ebinger:20} & 13.00 & 0.000e+00 &    0 & -1.37 &  0.14 \\ 
        PUSH &\cite{ebinger:20} & 14.00 & 0.000e+00 &    0 & -1.52 & -0.02 \\ 
        PUSH &\cite{ebinger:20} & 15.00 & 0.000e+00 &    0 & -0.27 & -0.28 \\ 
        PUSH &\cite{ebinger:20} & 16.00 & 0.000e+00 &    0 & -1.39 &  0.05 \\ 
        PUSH &\cite{ebinger:20} & 17.00 & 0.000e+00 &    0 & -1.52 & -0.35 \\ 
        PUSH &\cite{ebinger:20} & 18.00 & 0.000e+00 &    0 & -1.63 & -0.51 \\ 
        PUSH &\cite{ebinger:20} & 19.00 & 0.000e+00 &    0 & -1.06 &  0.11 \\ 
        PUSH &\cite{ebinger:20} & 20.00 & 0.000e+00 &    0 & -1.36 & -0.17 \\ 
        PUSH &\cite{ebinger:20} & 21.00 & 0.000e+00 &    0 & -0.76 &  0.54 \\ 
        PUSH &\cite{ebinger:20} & 22.00 & 0.000e+00 &    0 & -1.26 &  0.09 \\ 
        PUSH &\cite{ebinger:20} & 23.00 & 0.000e+00 &    0 & -0.79 &  0.30 \\ 
         NOM & \cite{nomoto:13} & 11.00 & 0.000e+00 &    0 & -1.76 & -2.00 \\ 
         NOM & \cite{nomoto:13} & 13.00 & 0.000e+00 &    0 & -1.84 & -2.43 \\ 
         NOM & \cite{nomoto:13} & 15.00 & 0.000e+00 &    0 & -1.54 & -2.05 \\ 
         NOM & \cite{nomoto:13} & 18.00 & 0.000e+00 &    0 & -1.51 & -2.27 \\ 
         NOM & \cite{nomoto:13} & 20.00 & 0.000e+00 &    0 & -1.09 & -1.59 \\ 
         NOM & \cite{nomoto:13} & 25.00 & 0.000e+00 &    0 & -0.79 & -1.11 \\ 
         NOM & \cite{nomoto:13} & 20.00 & 0.000e+00 &    0 & -1.43 & -1.77 \\ 
         NOM & \cite{nomoto:13} & 25.00 & 0.000e+00 &    0 & -1.18 & -1.66 \\ 
        \hline
    \end{longtable}
    \twocolumn

        Figure~\ref{fig:2} shows the distribution of [K/Mg] and [Sc/Mg] from several sets of CCSN yields, with progenitor initial mass between 11 and 25 \msun\ and initial metallicity between 0 and 0.05. As discussed in Sect. \ref{sec:obs}, some of the models shown in \figurename~\ref{fig:1} are not in \figurename~\ref{fig:2} and vice versa, because not all the sets provided both the \mco\ and \xc\ at the end of He burning and the explosive yields.
        Among the models used in \figurename~\ref{fig:1}, those from the  SIE, PGN, LIM, and BRI sets do not have C-O shell mergers, nor metallicity different than solar. Therefore, as discussed in the text, we did not include them in our comparison with observations of stars of low metallicity in \figurename~\ref{fig:3}. We added instead the NOM yield set of zero metallicity \citep{nomoto:13} and the PUSH set \citep{curtis:19,ebinger:20} of zero (z-series) and $\rm 10^{-4} Z_{\odot}$ metallicity (u-series). From the NOM set, we considered 6 CCSNe  from progenitors with initial masses between 11 and 25 \msun\, and 2 hypernova models with initial masses equal to 20 and 25 \msun. From the PUSH set we considered 12 models of the u-series, with initial masses between 11 and 25 \msun, and 13 models of the z-series, with initial masses between 11 and 23 \msun. The PUSH models only includes the composition of the inner part of the explosion which reaches temperatures above 1.75~GK. However, about 50$\%$ of the total Mg yield is produced outside of this regions and neither Sc nor K are significantly produced outside this region. Thus, we do not expect a significant impact on the [X/Mg] values. The production of the only stable isotope of Sc, \isotope[45]{Sc}, is a signature of proton-rich, neutrino-processed supernova ejecta \cite{froehlich:06,froehlich:06b} and recent multi-dimensional simulations have confirmed an enhancement of Sc compared to parameterized explosion models \cite{sieverding:20,wang:24}. While this enhancement of Sc could be up to a factor 10 \cite{sieverding:20}, it is not accompanied by an enhancement in the production of K. Therefore, we expect that the effects of multi-dimensional dynamics and neutrinos could improve the agreement with the observations, the effects alone are unlikely to be able to explain cases with an enhancement in both elements, Sc and K. Nonetheless, self-consistent simulations of metal-poor progenitor stars are needed to confirm the impact of multi-dimensional dynamics and neutrinos on the production of Sc and K. The models included in the comparison shown in \figurename~\ref{fig:3} are listed in \tablename~\ref{tab:kscmg}.

   \section{Observations} \label{subapp:fig4}

        \begin{figure}[!t]
            \includegraphics[width=0.48\textwidth]{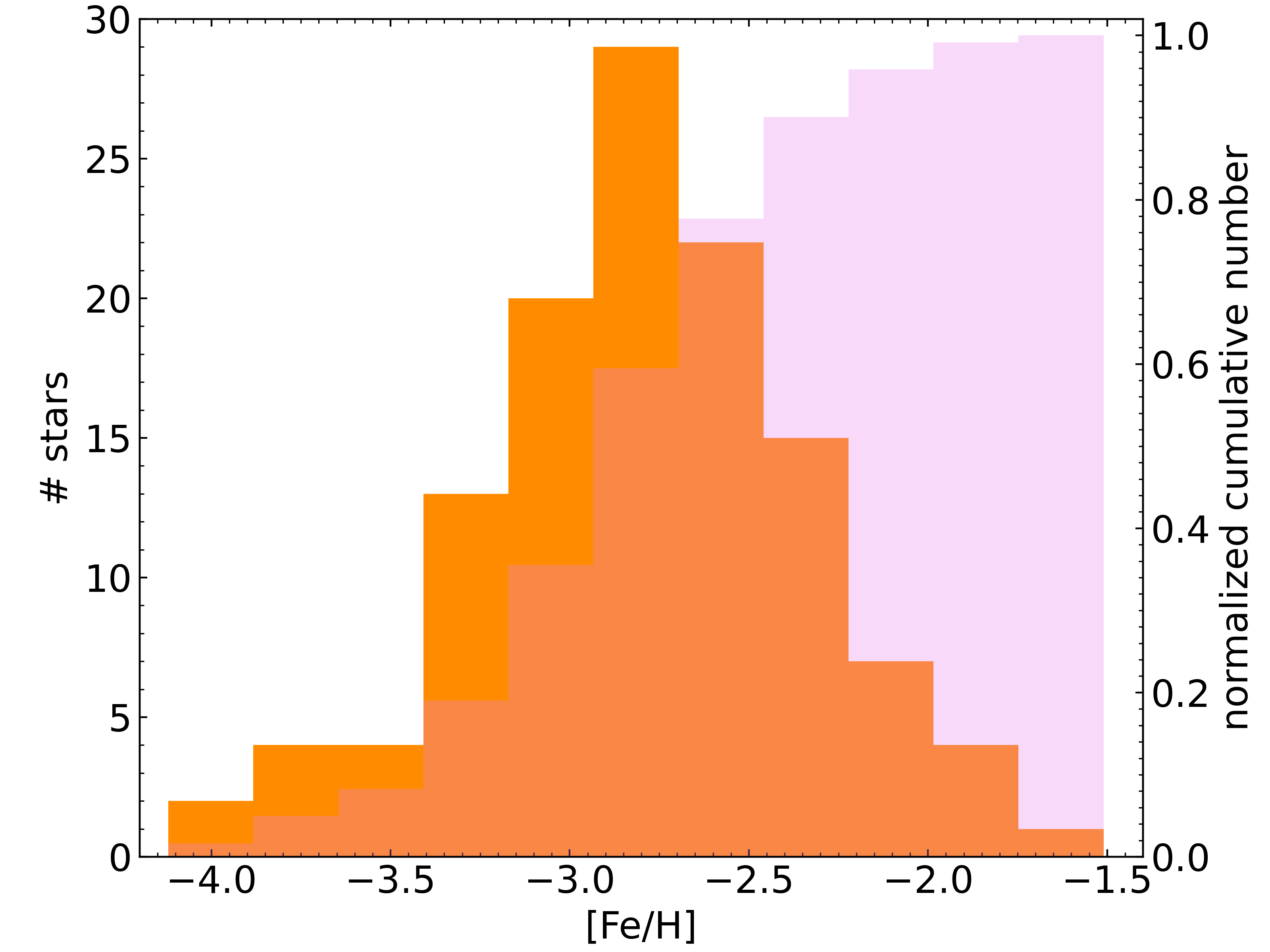}
            \caption{Metallicity histogram (orange bars) and normalized cumulative number of stars (pink bars) of the observational sample of 121 stars taken from JINAbase.}\label{fig:5}
        \end{figure} 

        In \figurename~\ref{fig:3} we compare the theoretical yield predictions with 37 observations of extremely metal poor (EMP, [Fe/H]$\leq-3$) stars taken from the JINAbase database \cite{jina} (as detailed in \tablename~\ref{tab:obs}). The full database contains 121 stars with [K/Mg] and [Sc/Mg] observations, with initial metallicity in the range $\rm -4.12 \leq [Fe/H] \leq -1.51$. More specifically, the data extracted from JINAbase is from \cite{roederer:10,cohen:13,roederer:14,siqueira:14,placco:14,placco:15,li:hn:15,casey:15,skuladottir:15,howes:15,howes:16}. A large number of the observations ($\simeq50\%$) are below [Fe/H]=--3, as indicated by the pink bars in \figurename~\ref{fig:5}, also showing that the distribution peaks at [Fe/H]=--2.8. The majority of these stars likely did not experience pollution from multiple generations of CCSNe, making them suitable candidates to establish the composition of the ejecta of single or few CCSNe. The average values for the full sample are $\rm \langle[K/Mg]\rangle = -0.006$ and $\rm \langle[Sc/Mg]\rangle = -0.410$,  while the subsample used for comparison with the models has slightly lower average values of $\rm \langle[K/Mg]\rangle = -0.047$ and $\rm \langle[Sc/Mg]\rangle = -0.523$. Since the JINAbase does not provide an error bar for the observed values, in \figurename~\ref{fig:3} we show a typical error bar from \cite{roederer:14}, the largest sample from which the data was extracted. With this error bar, we see that the EMP sample effectively represents the broader set of 121 stars. 

\onecolumn
\begin{table}
\caption{Observations used in \figurename~\ref{fig:3}.}
\label{tab:obs}
\centering
\begin{tabular}{llccc} 
\hline\hline
Simbad Identifier & Ref$^a$  & [Fe/H]  & [K/Mg]  & [Sc/Mg] \\
\hline
            $\rm 2MASS\_J00061720+1057419 $   & LI15a    & -3.26   &  0.07   & -0.65   \\
            $\rm 2MASS\_J01090507-0443211 $   & ROE14b   & -3.57   &  0.00   & -0.58   \\
            $\rm 2MASS\_J01265856+0135153 $   & LI15a    & -3.57   &  0.13   & -0.63   \\
            $\rm 2MASS\_J03173604-1517227 $   & ROE14b   & -3.06   &  0.04   & -0.47   \\
            $\rm 2MASS\_J04121388-1205050 $   & ROE14b   & -3.31   &  0.15   & -0.51   \\
            $\rm 2MASS\_J11272694-2352056 $   & ROE14b   & -3.36   & -0.29   & -0.61   \\
            $\rm 2MASS\_J15573010-2939228 $   & CAS15    & -3.02   &  0.11   & -1.13   \\
            $\rm 2MASS\_J19351908-6142243 $   & ROE14b   & -4.06   & -0.50   & -1.09   \\
            $\rm 2MASS\_J20070379-5834575 $   & ROE14b   & -3.32   &  0.10   & -0.42   \\
            $\rm 2MASS\_J20212839-1316336 $   & ROE14b   & -4.12   &  0.01   & -0.21   \\
            $\rm 2MASS\_J20525098-3419405 $   & ROE14b   & -3.31   & -0.28   & -0.39   \\
            $\rm 2MASS\_J21031185-6505088 $   & ROE14b   & -3.83   & -0.23   & -0.63   \\
            $\rm 2MASS\_J21035211-2942502 $   & ROE14b   & -3.87   & -1.33   & -1.83   \\
            $\rm 2MASS\_J21214583-3808579 $   & ROE14b   & -3.27   & -0.21   & -0.52   \\
            $\rm 2MASS\_J21570791-0434083 $   & ROE14b   & -3.13   & -0.27   & -0.39   \\
            $\rm 2MASS\_J22223599-0138275 $   & ROE14b   & -3.30   & -0.24   & -0.49   \\
            $\rm 2MASS\_J23325017-3359254 $   & ROE14b   & -3.11   & -0.27   & -0.42   \\
            $\rm 2MASS\_J23342669-2642140 $   & SIQ14    & -3.43   &  0.42   & -0.66   \\
            $\rm BD-13\_3695              $   & ROE14b   & -3.18   & -0.07   & -0.14   \\
            $\rm BD+38\_4955              $   & ROE14b   & -3.12   & -0.26   & -0.51   \\
            $\rm BPS\_CS\_22879-0012      $   & ROE14b   & -3.01   & -0.26   & -0.16   \\
            $\rm BPS\_CS\_22893-0010      $   & ROE14b   & -3.12   & -0.13   & -0.32   \\
            $\rm CD-24\_1782              $   & ROE14b   & -3.05   & -0.04   & -0.49   \\
            $\rm HD\_126587               $   & ROE14b   & -3.29   & -0.15   & -0.59   \\
            $\rm HD\_200654               $   & ROE14b   & -3.13   & -0.14   & -0.38   \\
            $\rm HE\_0048-6408            $   & PLA14a   & -3.75   &  0.15   & -0.60   \\
            $\rm HE\_0139-2826            $   & PLA14a   & -3.46   &  0.21   & -0.44   \\
            $\rm HE\_2233-4724            $   & PLA14a   & -3.65   &  0.26   & -0.44   \\
            $\rm Ross\_889                $   & ROE14b   & -3.09   & -0.05   & -0.13   \\
            $\rm SMSS\_J183719.09-262725.0$   & HOW15    & -3.18   & -0.06   & -0.29   \\
            $\rm TYC\_57-1052-1           $   & LI15a    & -3.36   &  0.07   & -0.77   \\
            $\rm UCAC4\_191-003009        $   & COH13    & -3.11   &  0.62   &  0.12   \\
            $\rm UCAC4\_242-000737        $   & COH13    & -3.18   & -0.04   & -0.34   \\
            $\rm UCAC4\_323-223299        $   & COH13    & -3.01   &  0.12   & -0.65   \\
            $\rm UCAC4\_363-001823        $   & COH13    & -3.34   &  0.70   & -0.53   \\
            \hline
\end{tabular}
\tablefoot{
$^a$ References: COH13 \cite{cohen:13}; ROE14b \cite{roederer:14}; SIQ14 \cite{siqueira:14}; 
PLA14a \cite{placco:14}; LI15a \cite{li:hn:15}; CAS15 \cite{casey:15}; HOW15 \cite{howes:15}.
}
\end{table}
\twocolumn

\end{appendix}

\end{document}